\documentclass[11pt]{article}
\pdfoutput=1
\usepackage{amsmath,amssymb,color,graphics,epsfig,mathrsfs}

\usepackage{amsfonts}

\newcommand{\hoch}[1]{$\, ^{#1}$}

\newcommand{\be}{\begin{equation}}
\newcommand{\ee}{\end{equation}}
\newcommand{\bea}{\setlength\arraycolsep{2pt} \begin{eqnarray}}
\newcommand{\eea}{\end{eqnarray}}

\def\ft#1#2{{\textstyle{\frac{\scriptstyle #1}{\scriptstyle #2} } }}
\def\fft#1#2{{\frac{#1}{#2}}}

\def\0{{\sst{(0)}}}
\def\1{{\sst{(1)}}}
\def\2{{\sst{(2)}}}
\def\3{{\sst{(3)}}}
\def\4{{\sst{(4)}}}
\def\5{{\sst{(5)}}}
\def\6{{\sst{(6)}}}
\def\7{{\sst{(7)}}}
\def\8{{\sst{(8)}}}
\def\sst#1{{\scriptscriptstyle #1}}

\def\ii{{\rm i}}

\begin{document}

\begin{flushright}
\hfill{KIAS-P12069}
\end{flushright}

\vspace{25pt}
\begin{center}
{\large {\bf Fermi Surfaces and Analytic Green's Functions from Conformal
Gravity}}

\vspace{10pt}
Jun Li\hoch{1}, Hai-Shan Liu\hoch{2}, H. L\"u\hoch{3} and Zhao-Long Wang\hoch{4}

\vspace{10pt}

\hoch{1} {\it Zhejiang Institute of Modern Physics, Zhejiang University, Hangzhou, 310027, China}

\vspace{10pt}

\hoch{2} {\it Institute for Advanced Physics \& Mathematics,\\
Zhejiang University of Technology, Hangzhou 310023, China}

\vspace{10pt}

\hoch{3}{\it Department of Physics, Beijing Normal University,
Beijing 100875, China}

\vspace{10pt}

\hoch{4} {\it School of Physics, Korea Institute for Advanced Study,
Seoul 130-722, Korea}

\vspace{40pt}

\underline{ABSTRACT}
\end{center}

We construct $T^2$-symmetric charged AdS black holes in conformal gravity. The most general solution up to an overall conformal factor contains three non-trivial parameters: the mass, electric charge and a quantity that can be identified as the massive spin-2 hair. We study the Dirac equation for the charged massless spinor in this background. The equation can be solved in terms of the general Heun's function for generic frequency $\omega$ and wave number $k$. This allows us to obtain the analytic Green's function $G(\omega, k)$ for both extremal and non-extremal black holes. For some special choice of back hole parameters, we find that the Green's function reduces to simpler hypergeometric or confluent hypergeometric functions.  We study the Fermi surfaces associated with the poles of the Green's function with vanishing $\omega$.  We find examples where the Fermi surfaces for non-Fermi liquids as well as the characteristic Fermi ones can arise. We illustrate the non-trivial differences in the Green's function and Fermi surfaces between the extremal and non-extremal black holes.


\thispagestyle{empty}

\pagebreak

\tableofcontents
\addtocontents{toc}{\protect\setcounter{tocdepth}{2}}


\newpage

\section{Introduction}

The AdS/CFT correspondence \cite{mald,gkp,wit} provides a new method of
studying strongly coupled field theories by constructing the
corresponding gravitational duals which are classical and relatively simpler. In particular, the Green's functions in the dual field theories can be derived
by studying the wave equations of the corresponding bulk fields such as the scalar or spinor fields. One of the most studied class of gravitational backgrounds is the charged black holes that are asymptotic to anti-de Sitter spacetimes (AdS) whose bulk vectors are dual to the currents in the boundary field theories.  Such bulk geometry allows one to investigate properties of some strongly coupled fermionic system at finite charge density such as non-Fermi liquids \cite{Lee:2008xf}. Analogous to those of the scalar operators, the fermionic Green's function can be obtained by analysing and solving the bulk Dirac equation.

The procedure of obtaining Fermi surfaces from the AdS/CFT correspondence was spelled out in detail in \cite{lmv,flmv}. (See also the review \cite{review}.) In general, for an electrically-charged black hole background, one can write down the Dirac equation for a charged spinor.  The mass of the spinor is typically chosen to be zero in practice except for analyzing the qualitative properties.  If one knows how to solve the equation, one can read off the Green's function in the momentum space from the asymptotic behaviour of the wave solution after imposing some appropriate horizon boundary condition.  The Green's function is expressed in terms of the black hole parameters and the electric charge $q$ of the spinor, as well as $\omega$, the frequency and $k$, the wave number. The poles of the Green's function indicate the existence of quasi particles.  A Fermi surface $k_F$ is defined as a pole of the Green's function in the momentum space with vanishing $\omega$.

Since the Fermi surfaces are determined by the $\omega=0$ Green's function, it is not necessary to know the Green's function for general $(\omega, k)$.   It appears sufficient to obtain wave solutions with $\omega=0$ and the associated Green's function $G(\omega=0,k)$.\footnote{As we shall see later, the situation is subtler for non-extremal backgrounds, where there is no criterium to impose the horizon boundary condition on the $\omega=0$ solutions alone.}  Nevertheless, it is of great interest to study the behaviour of the Green's function for small $\omega$ on or near the Fermi surface.
For extremal solution with zero temperature, a procedure was developed in \cite{lmv,flmv} to calculate such a Green's function. The Green's function for small $\omega$ can always be determined using a matching procedure provided that $G(0,k)$ is known. Near the Fermi surface, it takes the form \cite{lmv,flmv}
\begin{equation}
G(\omega,k)=-\fft{h_1}{(k-k_F)-v_F^{-1}\omega-h_2e^{{\rm i}\gamma_{k_F}}
\omega^{2\nu_{k_F}}}\,,\label{genfssmallw}
\end{equation}
where $h_1, v_F, h_2$ and $\nu_{k_F}$ are constants that can be determined.  In particular, $v_F$ here is the Fermi velocity.  It was known in the condensed matter physics that a Landau-Fermi surface of Fermi liquids is given by  $\nu_F=1$.  The Green's function with $\nu>\fft12$ shares the same characteristics of that near the Landau-Fermi surface. The situation with $\nu_F<\fft12$ is very different and the Green's function is associated with some non-Fermi liquids.

In practice, the Dirac equations typically cannot be solved analytically.  In \cite{lmv,flmv}, the extremal Reissner-Nordstr\"om AdS (RN) black holes in general dimensions were used as gravitational backgrounds.  It turns out that the Dirac equation cannot be solved even for $\omega=0$, except for $D=4$ \cite{d=4rncase}.  The above procedure were performed numerically.
Nevertheless, it shows that Fermi surfaces can arise from the extremal RN AdS black holes, and Green's functions for small $\omega$ near the Fermi surfaces indeed take the form (\ref{genfssmallw}).  Furthermore it was shown in \cite{lmv,flmv} that Fermi surfaces for non-Fermi liquids can also arise, which are characterised by $\nu_{F}<\fft12$.

Although a physical system can be adequately analysed numerically, it is more satisfying if one has analytic results. RN black holes in $D=5$ and $D=4$ are special cases of charged AdS black holes in gauged supergravities. The more general solutions are the three-charged \cite{bcs} and four-charged black holes \cite{duli} respectively, and their type IIB and M-theory embedding were given in \cite{10authors}. In \cite{gure}, the five-dimensional black hole was considered and reduced by setting two charges equal while turning off the third charge.  Taking a certain suitable extremal limit, Fermi surfaces of the resulting $G(0,k)$ can be read off analytically.  This enables one to construct an analytical expression for the constants $h_1, v_F, h_2$ and $\nu_{k_F}$ in (\ref{genfssmallw}). However, the extremal black hole suffers from a curvature singularity that is on the horizon.  The general $G(\omega,k)$ is unknown for this example.  Moreover, such a Green's function associated with a general non-extremal charged black hole is hitherto unknown. (Dirac fermions in non-zero temperature AdS RN black holes were first studied in \cite{Cubrovic:2009ye}.) On the other hand, the examples considered in \cite{lmv,flmv} and \cite{gure} are already among the simplest black holes in the usual two-derivative theories of gravity or supergravity. It is unlikely to find new black holes in these theories without exotic matter so that the Dirac equation becomes exactly solvable.

Charged black holes can also arise in conformal gravity.  In four dimensions, the Weyl-squared gravity is conformally invariant.  Its minimum coupling to the Maxwell field preserves the conformal symmetry.  Conformal gravity has attracted renewed interest recently.  It was shown that the Einstein-Weyl gravity has a critical point \cite{lpcritical} for which the seemingly inevitable ghost massive spin-2 excitations are replaced by modes that fall off logarithmically.   It was proposed that Einstein gravity can emerge from conformal gravity in the infrared region \cite{maldconf}.  Furthermore, conformal gravity can be supersymmetrised in the off-shell formalism \cite{ledu}.

The allowance of a vector field in conformal gravity implies that the theory should have a variety application in the AdS/CFT correspondence. It can provide the gravitational dual description of certain strongly interacting fermionic system at finite charge density, on which we focus in this paper. Furthermore, the system of (charged) massless Dirac spinor in the bulk is also conformally invariant, it is thus very natural to consider the minimally coupled spinor in conformal gravity.  The most general spherically-symmetric black holes carrying electric charges, up to an overall conformal factor, were obtained in \cite{Riegert:1984zz}.  For our purpose in this paper, we construct the analogous solution with the torus ($T^2$) topology.  We study the charged Dirac equation in this background and find that it can be solved exactly for general $(\omega, k)$.  This enables us to construct the general Green's function $G(\omega,k)$ for non-extremal black holes as well as their extremal limits. The results are expressed in terms of general Heun's functions, whose properties are less known.  We thus consider some special sub-classes of the black holes for which the wave solutions are reduced to hypergeometric functions.  This enables us to study the Green's functions in greater details, not only for extremal black holes but also for non-extremal ones.

The paper is organised as follows. In section 2, we review conformal gravity and the spherically-symmetric black holes. We then construct the static black holes with torus $T^2$ and hyperbolic $H^2$ horizons.  The most general solutions contain three non-trivial parameters, in addition to the cosmological constant. We study the global structure and analyse the thermodynamics.  We then consider two special subclasses of solutions with two parameters.  One was obtained previously in \cite{luwangsusylif} and it can be viewed as the pseudo-supersymmetric solution in the corresponding conformal off-shell supergravity.  The two-parameter solution has both inner and outer horizons. The solution is in general non-extremal, and it becomes extremal when the two horizons coalesce. The other class is the general extremal solution which has also two non-trivial parameters.  In section 3, we consider Dirac equations for a charged massless spinor in the black hole backgrounds and review the formalism of \cite{flmv}.  For the ``BPS'' black holes, the wave equation can be solved exactly.  We thus obtain the analytic Green's function for general $(\omega,k)$; it can be expressed in terms of hypergeometric functions.  We also obtain the Green's function in the extremal limit, and it turns out to be expressible in terms of confluent hypergeometric functions.  We study the Green's functions $G(0,k)$ and obtain the Fermi surfaces.  The results allow us to study the intrinsic differences on the boundary field theory between the extremal and non-extremal backgrounds.  We accomplish the above in section 4.

In section 5, we consider the two-parameter family of extremal black holes, and we find that $G(0,k)$ can be expressed in terms of hypergeometric functions.  This allows us to determine the Fermi surfaces.  In section 6, we demonstrate that the Dirac equation can be solved even in the most general black hole backgrounds.  However, the resulting Green's function $G(\omega, k)$ is expressed in terms of the general Heun's functions. We discuss how the result can be reduced to the previous simpler examples. Owing to the complication of the Heun's function, it is difficult to discuss many subtle properties of the Green's function; nevertheless, we are able to find many examples of Fermi surfaces.  The Green's function has a fascinating rich structure of spiked maxima in the $(\omega,k)$ plane. We conclude our paper in section 7.

\section{Charged black hole in conformal gravity}

\subsection{Review of conformal gravity}

Analytical charged AdS black holes also arise in conformal gravity in four dimensions. Conformal pure gravity is constructed from the Weyl-squared term.  Its conformal symmetry is preserved when it couples minimally to the Maxwell field.  The Lagrangian is given by
\begin{equation}
e^{-1}{\cal L} =  \ft12\alpha C^{\mu\nu\rho\sigma}
C_{\mu\nu\rho\sigma} + \ft13\alpha F^2\,,\label{conflag}
\end{equation}
where $e=\sqrt{-g}$, $F=dA$ and $C_{\mu\nu\rho\sigma}$ is the Weyl tensor. Note that the absolute value of the coupling of the Maxwell field can be arbitrary defined by a constant scaling of $A$. However its sign choice is non-trivial.  We have made this selection with the following two considerations.  The first is inspired by critical gravity \cite{lpcritical}.  If we add an cosmological Einstein Hilbert term  $e\, (R+6)$ to the Lagrangian, the critical point is precisely $\alpha=-\fft12$, in which case, the vector $A$ is non-ghost-like for our choice.  The second is the consideration that Einstein gravity may emerge from conformal gravity in the infrared limit \cite{maldconf}.  The conformal gravity with $\alpha=-\fft12$ can give rise to Einstein gravity $e\, (R+6)$.  Having determined the theory, it is straightforward to derive the equations of motion, given by
\begin{equation}
\nabla^{\mu} F_{\mu\nu}=0\,,\qquad
-\alpha (2\nabla^\rho\nabla^\sigma +
R^{\rho\sigma})C_{\mu\rho\sigma\nu} - \ft23\alpha (F_{\mu\nu}^2 -
\ft14 F^2 g_{\mu\nu})=0\,.
\end{equation}

It is worth pointing out that conformal gravity can be supresymmetrized in the off-shell formalism.  In ${\cal N}=1$, $D=4$ off-shell supergravity, the bosonic fields consist of the metric, a vector and a complex scalar $S + {\rm i} P$. The fermionic field involves only the off-shell gravitino $\psi_\mu$.  Up to and including the quadratic order in curvature, the theory allows four independent super-invariants. These comprise a ``cosmological term \cite{lpsw},'' the Einstein-Hilbert term \cite{sw,fv}, and two quadratic-curvature terms \cite{ledu}, one formed using the square of the Weyl tensor, and the other formed using the square of the Ricci scalar.  Supersymmetric solutions in these higher-derivative off-shell supergravities were obtained including Lifshitz, Schr\"odinger, gyratons \cite{luwangsusylif,liulu,lpgyr,llpp}.

The Lagrangian (\ref{conflag}) turns out to be the bosonic part of the super invariant, except that an analytical continuation of $A\rightarrow {\rm i}\, A$ was performed.  The resulting theory is called pseudo-supergravity with the Killing spinor equation \cite{luwangsusylif,llpp}:
\begin{equation}
\delta \psi_\mu=-D_{\mu} \epsilon + \ft{1}6 (2A_\mu - \Gamma_{\mu\nu} A^\nu)\Gamma_5
\epsilon - \ft16\Gamma_\mu (S + \Gamma_5
P)\epsilon=0\,.\label{susytrans}
\end{equation}

\subsection{General black holes and thermodynamics}

Up to an overall conformal factor, we find that conformal gravity admits the following static charged black hole solution,\footnote{As we shall see later, the wave equation of a massless spinor is invariant under the conformal transformation that is $r$-dependent only.  It follows that the Green's function and Fermi surfaces are independent of the $T^2$-symmetric conformal transformation.}
\begin{eqnarray}
ds^2 &=&-f dt^2 + \fft{dr^2}{f} + r^2
d\Omega_{2,\varepsilon}^2\,,\qquad A=-\fft{Q}{r} dt\,,\cr 
f&=&-\ft13\Lambda r^2 + c_1 r + c_0 + \fft{d}{r}\,,\qquad 3c_1 d
+\varepsilon^2 + Q^2 = c_0^2\,,\label{rieg}
\end{eqnarray}
where $\varepsilon=1,0,-1$ for $d\Omega_{2,\varepsilon}^2$ being the
metric for the unit sphere $S^2$ , torus $T^2$ and hyperbolic $H^2$.
The solution for $\varepsilon=1$ was given in \cite{Riegert:1984zz}.
Note that although (\ref{rieg}) contains the Schwarzschild black hole as a special solution, it does not contain the RN black hole.  This is important for this paper since the Dirac equation in the RN black hole is not analytically solvable.

    We now study the global structure and thermodynamics of the
general black hole.  (The thermodynamical properties for $Q=0$ were given in \cite{Lu:2012xu}.) Let $r_0>0$ denote the largest real root of $f(r)$.  The
metric on and outside the horizon runs from $r=r_0$ to the asymptotic AdS
boundary at $r=\infty$.  It is advantageous to express $Q$ in terms
of the remaining parameters and $r_0$, given by
\begin{equation}
Q=\sqrt{c_0^2  - \varepsilon^2  + \fft{3d^2}{r_0^2}+ \fft{3c_0
d}{r_0} - \Lambda r_0 d}\,.
\end{equation}
The temperature can be obtained by the standard method of
Euclideanising the metric and demanding the regularity on the horizon.
The entropy can be calculated directly from the Ward formula.  They are given by
\begin{equation}
T=-\fft{6d + 3 c_0 r_0 + \Lambda r_0^3}{12 \pi r_0^2}\,,\qquad
S=\ft16\alpha \varepsilon \omega_2- \fft{\alpha (3d + c_0
r_0)\omega_2}{6r_0}\,.
\end{equation}
The electric potential and charge are
\begin{equation}
\Phi = - \fft{Q}{r_0}\,,\qquad Q_e = \fft{\alpha \omega_2 Q}{12\pi}\,.
\end{equation}
As discussed in \cite{Lu:2012xu}, it is also necessary to introduce
a pair of thermodynamical quantities that can be viewed as the massive
spin-2 hair:
\begin{equation}
\Psi = \fft{\alpha (c_0-\varepsilon)\omega_2}{24\pi}\,,\qquad
\Xi=c_1\,.
\end{equation}
The free-energy can be obtained from the Euclidean action, given by
\begin{equation}
F=-\fft{\omega_2\alpha }{16\pi} \int_{r_0}^\infty r^2 dr \Big(\ft12
|{\rm Weyl}|^2 + \ft13 F^2\Big) = \fft{\alpha
\Big(2(c_0-\varepsilon)\varepsilon r_0 + (3\varepsilon-\Lambda
r_0^2)d\Big)}{24\pi r_0^2}\,.\label{freeenergy}
\end{equation}
Note that for the $T^2$ topology $\epsilon=0$, the sign of the free energy is determined by the sign of the parameter $d$. We can also treat the cosmological constant $\Lambda$ as a thermodynamical variable, since it is an integration constant in the solution. Its thermodynamical conjugate is given by
\begin{equation}
\Theta = \fft{\alpha \omega_2 d}{24\pi}\,.
\end{equation}
The energy calculation is more subtle.  It was shown \cite{Lu:2012xu} that both the ADT \cite{Abbott:1981ff,Deser:2002jk} and AMD \cite{Ashtekar:1984zz,Ashtekar:1999jx,Okuyama:2005fg,Pang:2011cs} methods would give rise to divergent results for this type of black holes. It can however be calculated using the method outlined in \cite{Lu:2012xu} by employing the Noether charge associated with the time-like Killing vector.  We find
\begin{equation}
E=\fft{\alpha (c_0 - \varepsilon) (\Lambda r_0^2-3c_0)}{72\pi r_0} +
\fft{\alpha (2 \Lambda r_0^2 -c_0 + \varepsilon)d}{24\pi r_0^2}\,.
\end{equation}
It is then straightforward to verify that the following
thermodynamical relations are satisfied:
\begin{equation}
dE = T dS + \Phi dQ_e + \Psi d\Xi + \Theta d\Lambda\,,\qquad F=E-TS
- \Phi Q_e\,.
\end{equation}
It is worth pointing out that the Smarr formula now becomes
\begin{equation}
E=2\Theta \Lambda + \Psi \Xi\,.
\end{equation}

\subsection{Comments on the massive spin-2 hair}

One characteristic of the black hole in conformal gravity is that there is a linear-$r$ term in the function $f$.  This term is sometime referred in the literature as the acceleration term, which is a misnomer.  Indeed, the analogous term appears in the Plebanski-Diemenienski solution that has a physical interpretation as the acceleration parameter.  However, in the Pebanski-Diemenienski solution, there is an overall conformal factor and the solution is stationary rather than static.  The asymptotic region of the Plebanski-Diemenienski solution is not at $r=\infty$, but where the conformal factor diverges.  A static solution like ours can have no acceleration, and the linear-$r$ term should be interpreted as the massive spin-2 hair associated with the thermodynamic pair $(\Psi,\Xi)$ discussed earlier.  If we consider linearized gravity around the AdS$_4$ vacuum, the spin-2 modes $h_{\mu\nu}$ in conformal gravity satisfy the following equation
\begin{equation}
(\Box + 2)(\Box + 4) h_{\mu\nu}=0\,.
\end{equation}
Here, without loss of generality, we set $\Lambda=-3$.  The first bracket gives rise to the (massless) graviton whilst the second bracket give rise to the
massive spin-2 mode with $M^2=-2$.  The solutions of these modes in AdS were obtained in \cite{Bergshoeff:2011ri} and the results were generalized to arbitrary dimensions and arbitrary integer spins \cite{Chen:2011in,Lu:2011qx}.  The falloff of a spin-2 mode is given by
\begin{equation}
h_{\mu\nu} \sim \fft{1}{r^{E-2}}\,,
\end{equation}
where $E(E-3)=M^2$.  Thus, although the massive spin-2 mode has a negative mass square, it satisfies the Breitenlohner-Freedman type of bound $M^2\ge -9/2$.  Indeed, for the massive spin-2 modes with $M^2=-2$, we have either $E=2$ or $E=1$, corresponding to the linear and constant ``falloffs''.  The effect of these modes appearing in the static black hole was discussed in the previous subsection.

It should be emphasized that although the black holes we considered in the previous subsection have terms with slower falloffs than the usual Schwarzschild AdS black holes, the asymptotic region is nevertheless the AdS$_4$ and hence the AdS/CFT correspondence applies in conformal gravity on the AdS$_4$ background.  The massive spin-2 modes are dual to some relevant spin-2 operators whose conformal dimensions can be read off from the asymptotic falloffs, namely $\Delta=2$ or $\Delta=1$, which are less than the dimensions of the boundary field theory.  The appearance of the massive spin-2 hairs in the black hole implies that the corresponding operators in the dual field theory take some non-vanishing expectation values.  The exact boundary conformal field theory is however unknown.  The purpose of this paper is to study the Green's function of the fermionic operator associated with the charged massless bulk spinor.

\subsection{A class of ``BPS'' black holes}

The general solution (\ref{rieg}) has a total of four parameters.  In this paper, we shall concentrate only on the solutions that are asymptotic to the AdS.  It follows that the parameter $\Lambda$ is trivial and can always be set to a fixed negative constant.  We shall let it be $\Lambda = -3$ so that the AdS has the unit radius. The general solution has then three non-trivial parameters, associated with the mass, charge and the massive spin-2 hair.

As discussed in section 2.1, conformal gravity can be supersymmetrised in the ${\cal N}=1$ off-shell formalism.  Although in our Lagrangian (\ref{conflag}) the vector field has changed from $A$ to ${\rm i} A$ compared to the supergravity theory, the resulting Killing spinor equation (\ref{susytrans}) is still valid. It was shown in \cite{luwangsusylif} that for the
static black hole configuration, the existence of a Killing spinor
requires that $\varepsilon=0$, {\it i.e.}~the topology of the black hole horizon is $T^2$.  Furthermore, we must have $d=0$ in
(\ref{rieg}).  Thus the ``BPS'' solutions take the form \cite{luwangsusylif}:
\begin{equation}
ds^2=-f dt^2 + \fft{dr^2}{f} + r^2 (dx^2 + dy^2)\,,\qquad A=a
dt\,,\label{genbulk0}
\end{equation}
where $f$, $\phi$ are functions of $r$ only:
\begin{equation}
f=r^2 - \mu r + Q\,,\qquad a=-\fft{Q}{r}\,,
\end{equation}
The solution has a local Killing spinor of (\ref{susytrans}) provided that
\begin{equation}
S=\fft{42 -5\mu r+ 6 r^2}{2r\sqrt{f}}\,,\qquad P=0\,.
\end{equation}
It is convenient to parameterise the solution as follows
\begin{equation}
f=(r-r_-)(r-r_+)\,,\qquad Q=r_- r_+\,,\qquad \mu=r_- +
r_+\,,\qquad r_+>0\,,\qquad r_+\ge r_-\,.\label{gensol1}
\end{equation}
Thus, in general the solution describes an asymptotic AdS charged
black hole with the outer horizon at $r=r_+$; $r=r_-$ is the inner horizon. The thermodynamical quantities are
\begin{eqnarray}
T\!\!\!\!&=&\!\!\!\!\fft{r_+-r_-}{4\pi}\,,\qquad S=-\ft16 \alpha r_-
r_+ \omega_2\,,\cr 
\Phi \!\!\!\!&=&\!\!\!\! r_-\,,\qquad Q_e=-\fft{\alpha r_- r_+
\omega_2}{12\pi} \,,\cr 
\Xi \!\!\!\!&=&\!\!\!\! r_- + r_+\,,\qquad \Psi=-\fft{\alpha r_- r_+
\omega_2}{24\pi} \,,\cr 
E \!\!\!\!&=&\!\!\!\! -\fft{\alpha r_- r_+ (r_- + r_+)
\omega_2}{24\pi}\,,\qquad F=0\,.\label{bpsthermo}
\end{eqnarray}
They satisfy the following thermodynamical relationship
\begin{equation}
dE=T dS + \Phi dQ_e + \Psi d\Xi\,,\qquad F=E - T S - \Phi Q_e\,.
\end{equation}
For this subclass of solutions, the free energy $F$ vanishes identically.  As we may see from the expression of the free energy (\ref{freeenergy}) that for $\epsilon=0$, the sign choice of the parameter $d$ dictates the sign of $F$.
Thus the above ``BPS'' black hole sits in the critical point of the Hawking-Page global phase transition.  The internal energy at this phase transition point is $E=TS + \Phi Q_e$.

     It is worth pointing out that although there exists a local
Killing spinor for general parameters, the function $S$ blows up on
the horizon $r=r_+$, implying that the solution is not truly supersymmetric.  However, this defect disappears in the extremal limit
with $r_+=r_-\equiv r_0$, giving
\begin{equation}
ds^2 = -(r-r_0)^2 dt^2 + \fft{dr^2}{(r-r_0)^2} + r^2 (dx^2 + dy^2)\,,\qquad
A=\fft{r_0(r-r_0)}{r} dt\,.\label{bpsext}
\end{equation}
In this limit, we have $T=0=F$ and
$E=\Phi Q_e$.  Note that if we require that $S$ be
constant, there exists an alternative extremal solution:
\begin{eqnarray}
ds^2= - \fft{(r-r_0)^6}{r^4} dt^2 + \fft{dr^2}{(r-r_0)^2} + r^2 (dx^2 + dy^2)\,,\qquad A= \fft{3r_0(r-r_0)^3}{r^3}\, dt\,.
\end{eqnarray}
Both extremal solutions run from the AdS$_2\times
T^2$ horizon at $r=r_0$ to AdS$_4$ at $r=\infty$.  It was shown that
these two extremal solutions are related one to another by an overall conformal
scaling \cite{luwangsusylif}.  {\it A priori}, they should be treated different solutions; however, as we shall see later, Fermi surfaces are invariant under the conformal transformations that preserve the $T^2$ isometry.

\subsection{Classes of non-BPS black holes}

The general solution (\ref{rieg}), which has three non-trivial parameters, is rather complicated for our purpose of studying the dual conformal field theory in some detail.  The previous ``BPS''  limit is a way of simplifying the solution, corresponding to setting $d=0$. We consider three more simplified cases with non-vanishing charge.  The first corresponds to $c_1=0$, the second corresponds to $c_0=0$ and the third is the general extremal solution.  In all these three cases, there are two non-trivial parameters, one of which is the electric charge.  For our purpose later, we shall consider only the $T^2$ topology, namely $\varepsilon=0$.  Without loss of generality, we set $\Lambda =-3$.

\bigskip
\noindent{\bf Case 1: $c_1=0$}
\medskip

Setting the parameter $c_1=0$ has the effect of turning off the thermodynamical quantity $\Xi$. The solution is given by
\begin{equation}
f=r^2 + Q + \fft{d}{r}\,,\qquad a=- \fft{Q}{r}\,.
\end{equation}
It admits an extremal limit, for which $f$ has a double zero at the horizon $r=r_0$. The parameters and the solution are given by
\begin{equation}
Q=-3r_0^2\,,\qquad d=2r_0^3\,,\qquad f=\fft{(r-r_0)^2 (r + 2 r_0)}{r}\,,\qquad a=-\fft{3r_0 (r-r_0)}{r}\,.\label{nbpsextr1}
\end{equation}
Note that we have made a gauge choice so that $a$ vanishes at $r=r_0$.   It is of interest to note that the same local metric also describe a non-extremal black hole when $r$ is negative.  (The horizon is at $r=-2r_0$ and the asymptotic AdS arises as $r\rightarrow -\infty$.)  Alternatively, we can take a view that non-extremal black holes arise when $r_0$ becomes negative.  Thus the solution describes a black  hole for any real and non-vanishing $r_0$.

\bigskip
\noindent{\bf Case 2: $c_0=0$}
\medskip

Setting the parameter $c_0=0$ has the effect of turning off the thermodynamical quantity $\Psi$. The solution is given by
\begin{equation}
f=r^2 + c_1 r - \fft{Q^2}{3c_1 r}\,,\qquad a=- \fft{Q}{r}\,.
\end{equation}
It also admits an extremal limit.  The corresponding parameters and the solution are given by
\begin{equation}
Q=-\ft32 r_0^2\,,\qquad c_1=-\ft32 r_0\,,\qquad
f=\fft{(r-r_0)^2 (2r+r_0)}{2r}\,,\qquad a=-\fft{3r_0(r-r_0)}{2r}\,.
\label{nbpsextr2}
\end{equation}
As usual, we have made a gauge choice so that $a$ vanishes at $r=r_0$.  As in the case 1,  the same local metric also describes a non-extremal black hole when $r$ is negative.  (The horizon is at $r=-r_0/2$ and the asymptotic AdS arises as $r\rightarrow -\infty$.)

\bigskip
\noindent{\bf Case 3: General extremal solution}
\medskip

It follows from (\ref{rieg}) that the function $f$ for the general extremal solution must take the form $f=(r-r_0)^2(r-r_1)/r$ with $r_0>0$ and $r_0>r_1$.
This implies that
\begin{equation}
c_0 = r_0 (r_0 + 2 r_1)\,,\qquad c_1 = -2 r_0 - r_1\,,\qquad
d = -r_0^2 r_1\,,\qquad Q = r_0 (r_0 - r_1)\,.
\end{equation}
This leads to the general extremal solution
\begin{equation}
f=\fft{(r-r_0)^2 (r-r_1)}{r}\,,\qquad a= \fft{(r-r_0)(r_0-r_1)}{r}\,.\label{genextr}
\end{equation}
(Note that if we let $r_1>r_0$, the solution becomes non-extremal black hole with horizon located at $r=r_1$.) If we set $r_1=0$, the solution becomes the ``BPS'' extremal solution (\ref{bpsext}).  If instead we let $r_1=2r_0$ or $r_1=\fft12 r_0$, we obtain the extremal solutions (\ref{nbpsextr1}) and (\ref{nbpsextr2}) respectively.

\section{Charged spinor and the Green's function}

In the previous section, we have constructed a variety of charged black holes in conformal gravity that are asymptotically AdS.  In this section, we review and discuss the formalism of deriving the Green's function from a charged spinor satisfying the Dirac equation \cite{lmv,flmv}.  The Dirac equation for a charged massless spinor $\psi$ in a generic gravitational background is given by
\begin{equation}
\gamma^\mu (\partial_\mu + \ft14\omega_{\mu}^{ab}\Gamma_{ab} - {\rm
i} q A_\mu)\Psi=0\,.\label{diracE0}
\end{equation}
Massive charged Dirac fermions were also considered in \cite{lmv,flmv}; We shall not consider them in this paper, since they break the conformal symmetry.  To derive the formalism, we consider a more general background in the form of
\begin{equation}
ds^2=-f dt^2 + \fft{dr^2}{h} + r^2 (dx^2 + dy^2)\,,\qquad A=a\,dt\,.
\label{genbulk}
\end{equation}
Note that the vector has a gauge symmetry $A\rightarrow A + d\Lambda$.  In the Dirac equation, this gauge symmetry has the effect of creating a phase factor in the fermion field $\psi$.  In particular, the constant shift in $A_t$ implies a constant shift in $\omega$. Thus in this paper, we shall always make a gauge choice such that $A$ vanishes on the horizon by an appropriate constant shift in $A_t$.  This gauge choice makes the boundary condition of the wave solution on the horizon easier to define.

The vielbein and the corresponding spin connection for (\ref{genbulk}) are given by
\begin{eqnarray}
&&e^0=f^{\ft12}\, dt\,,\qquad e^1 = r dx\,,\qquad e^2 = r dy\,,
\qquad e^3 =\fft{dr}{\sqrt{h}}\,,\cr 
&&\omega^0{}_3 = \ft12h^{\fft12} f' f^{-1} e^0\,,\qquad
\omega^{1}{}_3= h^{\fft12} r^{-1}
e^1\,,\qquad \omega^{2}{}_3= h^{\fft12} r^{-1} e^2\,.
\end{eqnarray}
The contribution of the spin connection to the Dirac equation
(\ref{diracE0}) can be absorbed by the scaling of the field
$\widetilde\Psi= (-g g^{rr})^{\fft14}\Psi$.  In general a wave function involves the frequency $\omega$ and the wave numbers $k_x$ and $k_y$.  Owing to the symmetry in the $(x,y)$ plane, we can set $k_y=0$ without loss of generality. In the Fourier mode
$\widetilde \Psi \sim e^{-{\rm i} \omega t + {\rm i} k x}\hat \Psi$
with the momentum lying in the $x$ direction only, the Dirac equation
becomes
\begin{equation}
\Big( -{\rm i} f^{-\fft12} (\omega + q \phi)\gamma^0 + h^{\fft12} \gamma^3
\partial_r + {\rm i} r^{-1} k \gamma^1\Big) \hat \Psi=0\,.\label{hatpsieom}
\end{equation}
Adopting the following conventions for the gamma matrices
\begin{gather}
\gamma^{0}=\begin{pmatrix}
{\rm i}\sigma_1 & 0\\
0 & {\rm i}\sigma_1
\end{pmatrix}\qquad
\gamma^{1}=\begin{pmatrix}
-\sigma_2 & 0\\
0 & \sigma_2
\end{pmatrix}\notag\qquad
\gamma^{2}=\begin{pmatrix}
0 & -\sigma_2\\
-\sigma_2 & 0
\end{pmatrix}\\
\gamma^{3}=\begin{pmatrix}
\sigma_3 & 0\\
0 & \sigma_3
\end{pmatrix}\notag
\qquad
\gamma^{5}=\begin{pmatrix}
0 & {\rm i}\sigma_2\\
-{\rm i}\sigma_2 & 0
\end{pmatrix},
\end{gather}
the Dirac equation reduces to two decoupled equations
\begin{eqnarray}
\bigl[{f}^{-\fft12}(\omega + q \phi)\sigma_1+ h^{\fft12}\sigma_3\partial_r
+(-1)^\alpha{r}^{-1}{\rm i}k \sigma_2\bigr]\hat \psi_\alpha=0\,,
\end{eqnarray}
where $\alpha=1,2$.  Note that only three gamma matrices $(\gamma^0,\gamma^3,\gamma^1)$ appear in the wave equation (\ref{hatpsieom}). The detail gamma matrix decomposition above is unnecessary except for the three. Thus, it is rather straightforward to generalise the formalism to arbitrary dimensions \cite{flmv}.

The equation for $\psi_2$ is related to the equation for $\psi_1$ by $k\rightarrow -k$.  Each $\psi_\alpha$ is a two-component spinor.
Let $\hat\psi_1=(u_1,u_2)^T$ and
$u_{\pm}=u_1\pm {\rm i} u_2$, we find that
\begin{eqnarray}
u_+'+\bar{\lambda}_1(r)u_+=\bar{\lambda}_2(r)u_-\,,\qquad
u_-'+\lambda_1(r)u_-=\lambda_2(r)u_+\,,\label{upmeom}
\end{eqnarray}
where
\begin{eqnarray}
\lambda_1(r)=\fft{{\rm i} (\omega+q a)}{\sqrt{f h}}\,,\qquad
\lambda_2(r)=-\fft{{\rm i} k}{r \sqrt{h}}\,.
\end{eqnarray}
(Note that if we had considered the massive Dirac equation, we would
have instead $\lambda_2(r)=\frac{m}{\sqrt{h}}-\fft{{\rm i} k}{r \sqrt{h}}$.) Here the bar denotes the complex conjugation. It follows that
\begin{align}
u_+''+\bar{p}_1(r)u_+'+\bar{p}_2(r)u_+ &=0\,,\label{up2ndeom}\\
u_-''+p_1(r)u_-'+p_2(r)u_- &=0\,,\label{um2ndeom}
\end{align}
where
\begin{equation}
p_1(r)=-\frac{\lambda_2'}{\lambda_2},\qquad
p_2(r)=|\lambda_1|^2-|\lambda_2|^2+p_1(r) \lambda_1+\lambda_1'.
\end{equation}
The advantage of writing the decoupled equations for $u_\pm$ instead
of $u_i$ is that these equations now involve only the metric
functions rather than their square roots.  In this paper, we opt to solve for $u_-$ first from the second-order differential equation (\ref{um2ndeom}).  We then obtain $u_+$ from (\ref{upmeom}), namely
\begin{equation}
u_+=\fft{u_-' + \lambda_1 u_-}{\lambda_2}\,.\label{upfromum}
\end{equation}
Note that although from (\ref{up2ndeom}) $u_+$ simply takes the form that is complex conjugate of $u_-$, the integration constants can only be determined through (\ref{upfromum}), so that the full set of solutions for $(u_+,u_-)$ have two independent integration constants rather than four.

As we shall see in the next section, for the ``BPS'' solutions (\ref{gensol1}), the wave functions $u_+$ and $u_-$ can be solved analytically for general $\omega$ and $k$, in terms of hypergeometric functions.  (The analysis on the general black hole is given in section 6.) By choosing the appropriate boundary condition on the horizon such that the waves are in-falling only, we can then read off the Green's function by studying the behaviour of the wave functions at AdS boundary, namely
\begin{equation}
G_1 = - G_2^{-1} = \lim_{r\rightarrow \infty} \fft{u_2}{u_1} =
-{\rm i}\lim_{r\rightarrow\infty}\frac{u_+-u_-}{u_++u_-}\,.\label{gfformula}
\end{equation}
Here, $G_1$ and $G_2$ are diagonal entries of the 2 by 2 matrix of the Green's function associated with the massless spinor \cite{lmv,flmv,gure}. In particular $G_1$ corresponds to the right-handed spinorial operator and the $G_2$ corresponds to the left, which we shall not consider in this paper.  In this ``standard'' quantization, the function $u_1$ is treated as source and $u_2$ is the response. It is clear that $u_1$ and $u_2$ are symmetric from the point of view of the wave function.  This leads to an alternative quantization in which the roles of $u_1$ and $u_2$ are reversed.  In this case, the Green's function of the right-handed spinor is given by $-G_2$ instead.

For a generic black hole of the form (\ref{genbulk}), if we could obtain the general wave solution and hence the Green's function, we could then determine the Fermi surfaces $k=k_F$ which are defined as poles of the Green's function of vanishing $\omega$. In practice, however, the Dirac equation cannot be solved exactly for generic $(\omega,k)$ in most backgrounds.  One can then try to solve the Dirac equation for simpler case with $\omega=0$.  This is enough for obtaining $G(0,k)$ and determining a Fermi surface.  It is still of interest to study the behaviour of the Green's function for small $\omega$ on or near the Fermi surface. In order to derive such properties on the Fermi surface, a procedure was developed in \cite{lmv,flmv} for the extremal RN black holes in general dimensions.  In fact the formalism applies for general extremal black holes, where the near-horizon geometry has an AdS$_2$ factor. This enables one to get the Green's functions in the small-$\omega$ limit as long as one knows the solution at $\omega=0$. Unfortunately, there was no analytic solution even for $\omega=0$,
except for $D=4$ \cite{d=4rncase}. The above procedure was only proceeded numerically.  Although the Green's function (\ref{genfssmallw}) can be reproduced numerically even for non-Fermi liquids, it is more satisfying to obtain some analytic expression for such a Green's function.

In \cite{gure}, an example of extremal background was discussed in which $G(0,k)$ and hence the Fermi surfaces could be obtained exactly. This gives rise to an analytical expression for the Green's function of small $\omega$, namely (\ref{genfssmallw}).  However, the general expression of $G(\omega,k)$ is still unknown.  In addition, an analytic expression of the Green's function for a non-extremal black hole is also lacking in the literature.  This story changes if we use instead the black hole backgrounds of conformal gravity: the Dirac equations become exactly solvable!  We shall proceed in the next section.

To end this section, we would like to comment on the effect of the conformal symmetry on the Fermi surfaces.  If a conformal transformation does not alter the global structure such as the asymptotic infinity and the horizon, the resulting new solution can be viewed as equivalent to the original one.  Since the Dirac equation is covariant under the conformal transformation, the Green's function $G(\omega,k)$ and hence the Fermi surfaces are invariant under the  conformal transformations.

\section{Exact Green's function from the ``BPS'' black holes}

As we shall demonstrate in section 6, the Dirac equation (\ref{diracE0}) on the general $T^2$-symmetric background (\ref{rieg}) with $\varepsilon=0$ can be solved exactly, and the analytic Green's function for general $(\omega,k)$ can be obtained.  However, the results are expressed in terms of the general Heun's functions, which are less studied. In fact there is no Heun's function defined in Mathematica, and very limited properties of these functions are coded in Maple. In this section, we study the Dirac equation in the simpler ``BPS'' backgrounds, given in section 2.2,  namely, the black holes given by (\ref{genbulk0}) with (\ref{gensol1}).  We make a gauge choice
\begin{equation}
A=(r_--\fft{r_+ r_-}{r}) dt\,,
\end{equation}
so that it vanishes on the horizon $r=r_+$.

\subsection{Green's function from non-extremal black holes}
Let us first consider the general solution (\ref{gensol1}) with two parameters $(r_+,r_-)$. We have
\begin{equation}
\lambda_1(r)={\rm i} \fft{ (\omega+q r_-)r -q r_+r_-}{r (r - r_+)(r-r_-)}\,,\qquad \lambda_2(r)=-\fft{{\rm i} k}{r\sqrt{(r - r_+)(r-r_-)}}\,,
\end{equation}
It turns out that the wave equations for $u_+$ and $u_-$ are exactly solvable.
We organize the solutions as in-falling and outgoing modes on the horizon. The in-falling solutions are given by
\begin{eqnarray}
u_+^{\rm in} &=& -\fft{2k}{4\omega + {\rm i} (r_+-r_-)} r^{-1 + {\rm i} q} (r-r_+)^{\fft12 - {\rm i} \Omega} (r-r_-)^{ \fft12 - {\rm i} q + {\rm i} \Omega}
\times\cr
&& {}_2F_1\Big[1 - {\rm i} q + \nu,\, 1-{\rm i} q-\nu;\, \ft32 - 2 {\rm i} \Omega;\,-\fft{r_-(r-r_-)}{r(r_+-r_-)}\Big]\,,\cr
u_-^{\rm in} &=& r^{{\rm i} q} (r-r_+)^{-{\rm i} \Omega} (r-r_-)^{-{\rm i} q + {\rm i} \Omega}\times\cr
&& {}_2F_1\Big[-{\rm i} q + \nu,\, - {\rm i} q - \nu;\, \ft12 -2{\rm i} \Omega;\, -\fft{r_-(r-r_+)}{r(r_+-r_-)}\Big]\,,\label{rpminfalling}
\end{eqnarray}
where
\begin{equation}
\Omega = \fft{\omega}{r_+ - r_-}=\fft{\omega}{4\pi T}\,,\qquad \nu =\sqrt{\fft{k^2}{r_+ r_-} - q^2}=\sqrt{\fft{k^2}{Q} - q^2}\,.
\label{omeganu}
\end{equation}
Note that $T$ is the temperature of the background black hole, given in (\ref{bpsthermo}). These solutions are called in-falling since near the horizon, the wave function behaves like
\begin{equation}
\psi \sim \exp\Big( -{\rm i} \omega t + {\rm i} \Omega \log(\fft{r-r_+}{4\pi T})\Big)\,.\label{rpmpsinearhorizon}
\end{equation}
The outgoing wave functions are given by
\begin{equation}
u_+^{\rm out} \sim (u_-^{\rm in})^*\,,\qquad
u_-^{\rm out} \sim (u_+^{\rm in})^*\,,
\end{equation}
where $*$ denotes the complex conjugate.  The outgoing solution should be discarded for the black hole background.
To be precise, the outgoing solution gives rise to the advanced Green's function which we shall not consider in this paper.

Following the discussion in section 3, we find that the Green's function for general $\omega$ and $k$ is given by
\begin{equation}
G(\omega, k) = {\rm i} \fft{1 - 2{\rm i} (\gamma+ 2\Omega)}{1 + 2{\rm i} ( \gamma-2\Omega)}\,,\label{bpsgreen1}
\end{equation}
where
\begin{eqnarray}
\gamma &=& \fft{{}_2F_1 [1-\nu - {\rm i} q,\, 1 + \nu - {\rm i} q;\,
\ft32 + {\rm i} \Omega;\, -\fft{r_-}{r_+ - r_-}]\,k}{{}_2F_1[
-\nu - {\rm i} q,\, \nu - {\rm i} q;\, \ft12 + {\rm i} \Omega;\,
-\fft{r_-}{r_+-r_-}]\,(r_+-r_-)}\cr
&=& \fft{{}_2F_1 [1-\nu - {\rm i} q,\, 1 + \nu - {\rm i} q;\,
\ft32 + {\rm i} \Omega;\, -\fft{r_-}{4\pi T}]\,k}{{}_2F_1[
-\nu - {\rm i} q,\, \nu - {\rm i} q;\, \ft12 + {\rm i} \Omega;\,
-\fft{r_-}{4\pi T}]\,(4\pi T)}\,,\label{bpsgreen2}
\end{eqnarray}
The Green's function for $\omega=0$ is then simply given by (\ref{bpsgreen1})
and (\ref{bpsgreen2}) with $\Omega$ set to zero.  It is clear that the general Green's function $G(\omega,k)$ is complex and it remains complex for $\omega=0$.  By contrast, as we shall see in the next subsection, $G(0,k)$ for extremal black hole becomes real.  Interestingly, in non-extremal backgrounds, the quantity $\nu$ does not have any particular physical importance and there is no requirement that it be real. The situation is changed in the extremal limit, where $\nu$ has to be real for stability.

We find that for some special parameter choices, the hypergeometric functions in the Green's function degenerate. For example, if we let $q=0=\omega$, we have
\begin{equation}
G={\rm i} \fft{ {\rm i} + \tanh \Big(\fft{2k}{\sqrt{r_+r_-}} {\rm arcsinh} \fft{r_-}{r_+-r_-}\Big)}{{\rm i} - \tanh \Big(\fft{2k}{\sqrt{r_+r_-}} {\rm arcsinh} \fft{r_-}{r_+-r_-}\Big)}\,.
\end{equation}
For $\nu=\ft14$ and $\omega=0$, we have
\begin{equation}
G={\rm i}\fft{2 - {\rm i}\sqrt{(1 + 16 q^2)r_+r_-}\, \gamma}{
   2 + {\rm i}\sqrt{(1 + 16 q^2)r_+r_-}\, \gamma}
\end{equation}
where
\begin{equation}
\gamma = - \fft{2{\rm i}}{(1-4{\rm i} q) \sqrt{r_- (r_+-r_-)}}\fft{\Big(1 - {\rm i} \sqrt{\fft{r_-}{r_+-r_-}}\Big)^{-\fft12 + 2 {\rm i} q} -\Big(1 + {\rm i} \sqrt{\fft{r_-}{r_+-r_-}}\Big)^{-\fft12 + 2 {\rm i} q}}{\Big(1 - {\rm i} \sqrt{\fft{r_-}{r_+-r_-}}\Big)^{\fft12 + 2 {\rm i} q} + \Big(1 + {\rm i} \sqrt{\fft{r_-}{r_+-r_-}}\Big)^{\fft12 + 2 {\rm i} q}}\,.
\end{equation}

\subsection{Green's function from the extremal black hole}

    We now consider the extremal case with $r_-=r_+\equiv r_0$. This case was
discussed in some detail in \cite{lwshort}. Although the results in this case can be obtained by taking a subtle extremal limit on the hypergeometric functions in the previous subsection, (further discussion is given in section 4.6,)  it is of interest to study the extremal case on its own and we shall discuss this example independently. In this case, we have
\begin{equation}
\lambda_1 = {\rm i} \fft{(\omega+ q r_0) r - qr_0^2}{r(r-r_0)^2}\,,\qquad
\lambda_2 = - \fft{{\rm i}k}{r(r-r_0)}\,.
\end{equation}
The in-falling solutions of $u_\pm$ can be obtained, given by
\begin{eqnarray}
u_-(r)=z^{-\fft12} W_{\kappa,\nu}(z)\,,\qquad
u_+(r)=\fft{{\rm i} k}{r_0} z^{-\fft12} W_{-\kappa^*,\nu}(z)\,,\label{upmsol}
\end{eqnarray}
where $W_{\kappa,\nu}(z)$ denotes the Whittaker function. The new notations above are specified as follows
\begin{equation}
\kappa = \ft12 + {\rm i} q\,,\qquad \nu = \sqrt{\ft{k^2}{r_0^2} - q^2}\,,\qquad
z=-\fft{2{\rm i} \omega\, r}{r_0(r-r_0)}\,,\label{extrknz}
\end{equation}
and $\kappa^*$ is the complex conjugate of $\kappa$.  As we shall discuss at the beginning of subsection 4.4, the quantity $\nu$ has to be real in this extremal limit, which implies that $|k|\ge |q r_0|$. The sign choice of $\nu$ is however conventional, and it has an effect on defining which mode is in-falling. The following identities are useful
\begin{eqnarray}
zW'_{\kappa,\nu}(z) &=& (\kappa - \ft12 z) W_{\kappa,\nu} (z) -\left(\nu^2- (\kappa-\ft12)^2\right) W_{\kappa -1,\nu} (z)\cr
&=& -(\kappa -\ft12 z)W_{\kappa,\nu} - W_{\kappa +1,\nu} (z)\,.
\end{eqnarray}
As $r$ approaches the horizon $r=r_0$, we have $z\rightarrow \infty$, and the Whittaker function behaves as
\begin{equation}
W_{\kappa,\nu} \rightarrow e^{-\fft12 z} z^{\kappa} (1 + {\cal O}(z^{-1}))\,.
\end{equation}
It is then straightforward to recognise that $\psi$ is in-falling only.  The retarded Green's function, associated with the in-falling modes, can now be read off straightforwardly, given by
\begin{equation}
G_1=-G_2^{-1} = {\rm i} \fft{1-\gamma}{
1  +\gamma}\,,
\end{equation}
where
\begin{equation}
\gamma = \fft{{\rm i} k\,U (1-{\rm i} q + \nu, 1 + 2\nu, -\fft{2{\rm i}\omega}{r_0})}{
r_0\, U(-{\rm i} q + \nu, 1 + 2\nu, -\fft{2{\rm i}\omega}{r_0})}\,.\label{extrgamma}
\end{equation}
Here we have used the identity
\begin{equation}
W_{\kappa, \nu} (z) = \exp(-\ft12 z) z^{\nu + \fft12} U(\nu-k+ \ft12, 1 + 2\nu; z)\,,
\end{equation}
where $U$ is the confluent hypergeometric function of the second kind, defined by
\begin{equation}
U(a,b,z) = \fft{\Gamma(1-b)}{\Gamma(a-b+1)} {}_1F_1 (a;b;z) +
\fft{\Gamma(b-1)}{\Gamma(a)} z^{1-b} {}_1F_1 (a-b+1;2-b;z)\,.\label{hyperu}
\end{equation}

We can also obtain the result directly from (\ref{bpsgreen1}) and (\ref{bpsgreen2}) by taking a limit of $r_-\rightarrow r_+\equiv r_0$.  Since the last argument in the hypergeometric function in (\ref{bpsgreen2}) blows in this limit, it is convenient to use the following identity
\begin{eqnarray}
{}_2F_1(a,b;c;z)&=&
\frac{\Gamma(c)\Gamma(b-a)}{\Gamma(b)\Gamma(c-a)}
(-z)^{-a}{}_2F_1(a,1+a-c;1+a-b;z^{-1})\cr
&&+\frac{\Gamma(c)\Gamma(a-b)}{\Gamma(a)
\Gamma(c-b)}(-z)^{-b}{}_2F_1
(b,1+b-c;1+b-a;z^{-1})\,,
\end{eqnarray}
for $z\not\in (0,1)$.

    It is worth emphasizing an important difference between the wave
functions in the extremal and non-extremal black holes.  In the extremal background, the general wave solution for the $u_-$ with $\omega=0$ is reduced to
\begin{equation}
u_- = c_1 \Big(1 - \fft{r_0}{r}\Big)^\nu + c_2 \Big(1 - \fft{r_0}{r}\Big)^{-\nu}\,.\label{extremalomega0}
\end{equation}
($u_+$ is analogous.) This implies that for $\nu>0$, the wave function is well defined on the horizon only when $c_2=0$.  This allows one to pick the correct mode even when we do not have the more general solution with non-vanishing $\omega$, in which case, the concept of in-falling and outgoing becomes irrelevant.  This implies that in the extremal case, the Green's function can be determined without any knowledge with $\omega\ne 0$.  Furthermore, when the near-horizon geometry has AdS$_2\times T^2$, the matching procedure of \cite{flmv} is always applicable with the same expression of the self-energy term. This enables one to get the Green's functions in the small-$\omega$ limit as long as one knows the solution at $\omega=0$. The situation is very different in the general non-extremal solution.  As we can saw in section 4.1, when $\omega=0$, both solutions of $u_-$ are well defined on the horizon.  We are able to select the right mode only because we have the general solution with non-vanishing $\omega$, which allows us to separate the in-falling mode from the outgoing mode.  If we can solve the Dirac equation only for $\omega=0$ in the non-extremal background, we shall not be able to fix the boundary condition of the $\omega=0$ solution alone on the horizon.  Further distinguishing features of extremal and non-extremal backgrounds will be addressed in subsections 4.4, 4.5 and 4.6.

\subsection{Fermi surface from the extremal black hole}

Having obtained the Green's function in the momentum space for general $\omega$ and $k$, we can now study the Fermi surfaces. They are characterised by the poles of the Green's function for vanishing $\omega$.  Since we have ${}_1F_1[a,b,0]=1$ for generic $a$ and $b$, it follows that for any positive $\nu$, the second term in the confluent hypergeometric function (\ref{hyperu}) dominates.  Thus in the $\omega\rightarrow 0$ limit, we have
\begin{equation}
G(0, k)={\rm i} + \fft{2k}{\nu r_0 + {\rm i} (k- q r_0)}=\sqrt{\fft{k+ q r_0}{k- q r_0}}\,,\label{extrw0}
\end{equation}
which is a real quantity provided that $\nu$ is real.  Note that for half integer $\nu$, the identity (\ref{hyperu}) diverges. However, the ratio $\gamma$  (\ref{extrgamma}) remains finite and the resulting Green's function takes the same form as the above. Thus for a given extremal black hole with $r_0$ and the given charge $q$ of the spinor, there is only one Fermi surface:  $k= q r_0>0$ for the standard quantization or $k=- q r_0<0$ for the alternative quantization, both correspond to the vanishing of $\nu$.

     Since the Fermi surface in this case occurs at $\nu=0$ only,
there is a delicate limiting procedure for investigating the Green's function at the vicinity of the Fermi surface.  Let us first examine the standard quantization case whose Fermi surface occurs at $k=k_F\equiv q r_0$.  If $\omega/r_0$ approaches zero much faster than $(k-k_F)$, we find that
\begin{equation}
G_1 \sim \fft{\sqrt{2 k_F (k-k_F)}}{(k-k_F) + q \sqrt{\fft{2(k-k_F)}{k_F}}\, \omega +
2(k-k_F)(-\fft{2{\rm i}\, \omega}{r_0})^{\sqrt{\fft{2(k-k_F)}{k_F}}}}\,.
\end{equation}
Comparing with the general formula of the Green's function near the Fermi surface (\ref{genfssmallw}), we find that
\begin{eqnarray}
v_F^{-1}&= &0-q\sqrt{\frac{2(k-k_F)}{k_F}}+\cdots\,,\cr
h_1&=&0-\sqrt{2k_F(k-k_F)}+\cdots\,,\qquad h_2=0-2(k-k_F)+\cdots.
\end{eqnarray}

    For the alternative quantization, the Fermi surface occurs at $k=k_F\equiv
-qr_0<0$.  Again in the limit where $\omega/r_0$ approaches zero much faster than $(k-k_F)$, we have
\begin{equation}
G_2 =-\fft{\sqrt{2 k_F (k-k_F)}}{(k-k_F) - q \sqrt{\fft{2(k-k_F)}{k_F}}\, \omega +
2(k-k_F)(-\fft{2{\rm i}\, \omega}{r_0})^{\sqrt{\fft{2(k-k_F)}{k_F}}}}\,.
\end{equation}
Comparing with the general formula (\ref{genfssmallw}), we have
\begin{eqnarray}
v_F^{-1}&=&0+q\sqrt{\frac{2(k-k_F)}{k_F}}+\cdots\,,\cr
h_1&=&0+\sqrt{2k_F(k-k_F)}+\cdots\,,\qquad h_2=0-2(k-k_F)+\cdots.
\end{eqnarray}

It is instructive also to consider how the Green's function behaves with small $\omega$, after we have literally fixed $k=k_F$.  We find
\begin{eqnarray}
k=q r_0:&& G_1 = -{\rm i} + 2 q \Big(\log(-\fft{2{\rm i} \omega}{r_0}) -2\gamma -\chi({\rm i} q)\Big)\,,\cr
k=-q r_0:&& G_2 = -{\rm i} + 2 q \Big(\log(-\fft{2{\rm i} \omega}{r_0}) -2\gamma -\chi({\rm i} q)\Big)\,,
\end{eqnarray}
where $\gamma$ is the Euler number and $\chi$ is the digamma function.
In both cases, the divergences are logarithmic in small $\omega$.  This is different from the general formula (\ref{genfssmallw}) with $k=k_F$.

The $\nu=0$ Fermi surface describes an extreme situation of some Non-Fermi liquids.
It was shown to exist in the extremal RN black hole \cite{lmv,flmv}.  As we shall see later, it can also arise in the more general two-parameter extremal solutions. The example discussed in this subsection is somewhat unusual in that this is the only Fermi surface.

\subsection{Imaginary $\nu$ and Fermi surfaces from non-extremal black holes}

In the above discussion of extremal background, we have considered the case where $\nu$, whose expression is given by (\ref{extrknz}), is a real number.  The general expression of $\nu$ for the non-extremal solutions is given by (\ref{omeganu}). Naively, one would expect that this quantity continues to be real, in which case, we find that there is no sign of Fermi surfaces in the non-extremal background.

We now consider the possibility that $\nu$ is pure imaginary.  There are two situations that this can arise.  The first is when $r_-$ is also positive.  In this case, $\nu$ becomes imaginary if we have
\begin{equation}
q^2 > \fft{k^2}{r_+ r_-}\,.
\end{equation}
The second situation is when $r_-$ is negative.  In this case, $\nu$ is pure imaginary for any $(k,q)$ that do not vanish simultaneously.  In either case, we shall define $\nu = {\rm i}\mu$ where
\begin{equation}
\mu = \sqrt{q^2 - \fft{k^2}{r_+ r_-}}\,.
\end{equation}

     For the extremal black hole, the solution for the wave function $u_-$ with
$\omega =0$ is given by
\begin{eqnarray}
u_- &=& c_1 \Big(1 - \fft{r_0}{r}\Big)^\nu + c_2 \Big(1 - \fft{r_0}{r}\Big)^{-\nu}\cr
&=& \Big(1 - \fft{r_0}{r}\Big)^{{\rm i}\mu} + c_2 \Big(1 - \fft{r_0}{r}\Big)^{-{\rm i}\mu}\,.\label{infiniteosc}
\end{eqnarray}
Thus, we see that if $\nu$ is real and positive, the wave function is well-defined on the horizon by setting $c_2=0$.  This is in fact exactly the near-horizon behaviour of the general in-falling solution (\ref{upmsol}) in the $\omega\rightarrow 0$ limit.  If on the other hand, $\nu={\rm i}\mu$ is pure imaginary, the wave function becomes oscillatory on the horizon.  The oscillatory mode with $\omega=0$ on the horizon makes it impossible to fix the horizon boundary condition. It was argued that this corresponds to an infinite number of particle creation and suggests that the geometry becomes unstable \cite{flmv}.  Further discussion will be given later.

This instability does not occur in the non-extremal solutions.  It is clear
that the hypergeometric functions in (\ref{rpminfalling}) are well behaved when $r\rightarrow r_+$.  It follows that the fermionic wave function in the near-horizon behaves simply as (\ref{rpmpsinearhorizon}) and there is no oscillatory behaviour when $\omega=0$. Thus as long as the black hole is non-extremal, imaginary $\nu$ does not imply instability and hence should be considered. It is worth drawing attention that for an extremal black hole, there usually exists a decoupling limit in which the metric becomes its near-horizon geometry AdS$_2\times T^2$.  The quantity $\nu$ has a clear physical interpretation that it measures the conformal weights of the dual operators in the boundary field theory of the AdS$_2$.  This implies that $\nu$ must be real.  Such a decoupling limit is absent in non-extremal solutions and $\nu$ has no apparent physical significance and it could be real or imaginary.  In fact, as we see in section 6, in the wave solution and the Green's function for the most general black holes, there is no parameter combination that can be recognised as $\nu$ at all.  Only in the extremal limit, such a combination appears.  The appearance of the $\nu$-like quantity in this non-extremal case is accidental. As we have mentioned, there is no Fermi surface at all for real $\nu$ in non-extremal solutions.  Interestingly, Fermi surfaces emerge quite frequently when $\nu$ is taken to be pure imaginary.  In other words, there is a restriction on the parameter space $(k,q)$ such as $|k|\ge |q r_0|$ in the extremal black hole, but such a restriction disappears in the non-extremal background.

     For $\omega=0$ and $\nu={\rm i} \mu$, the Green's function is given by
\begin{equation}
G(0,k)= {\rm i} \fft{1 - 2{\rm i}\gamma}{1 + 2{\rm i}\gamma}\,,\qquad
\gamma = \fft{{}_2F_1 [1-{\rm i}(\mu+q),\, 1 + {\rm i}(\mu-q);\,
\fft32;\, \fft{r_-}{r_- - r_+}]\,k}{{}_2F_1[
-{\rm i}(\mu + q),\, {\rm i}(\mu - q);\, \fft12;\,
\fft{r_-}{r_--r_+}]\,(r_+-r_-)}\,.\label{mugreen}
\end{equation}
Note an important distinguishing feature that the Green's function $G(0,k)$ above is not real, whilst it is always real in the extremal limit.  This is related to the fact that in the non-extremal case, both wave functions are well defined on the horizon when $\omega=0$, and the wave functions $u_+$ and $u_-$ are organized in terms as in-falling or outgoing modes.  This implies that $u_+$ and $u_-$ are not complex conjugate to each other, and the resulting Green's function is therefore not real, even when $\omega=0$.

Fermi surfaces can then emerge when $G(0,k)$ diverges or vanishes (for alternative quantization) for certain choice of $(k,q)$.  Since $G(0,k)$ is complex, it is unlikely that it diverges literally, since it would demand the vanishing of the real and complex parts of the inverse of the Green's function simultaneously.  Our explicit examples however demonstrate that there exist values of $k$ for which $|G(0,k)|$ becomes large in many orders of magnitude.  These can be regarded as Fermi surfaces in all practical purpose. Since an analytical expression for such conditions of $(k,q)$ is unlikely to exist, we shall present two illustrative examples. In both examples we examine the denominator of the Green's function (\ref{mugreen}), namely
\begin{equation}
{\cal D}=1 + 2 {\rm i}\gamma\,.
\end{equation}
The ``zeros'' of ${\cal D}$ signal the Fermi surfaces.  Note that in general ${\cal D}$ is a complex number.  For a given black hole with $r_\pm$ fixed, and given $q$, we have only one real variable $k$ to vary. It is hence rather unlikely that both the real and imaginary parts of ${\cal D}$ approach zero simultaneously for certain values of $k$.  We are looking for surfaces for which the minimum of $|1 + 2{\rm i} \gamma|$ is smaller in many orders of magnitude.

\begin{figure}[ht]
\ \ \ \ \ \includegraphics[width=7cm]{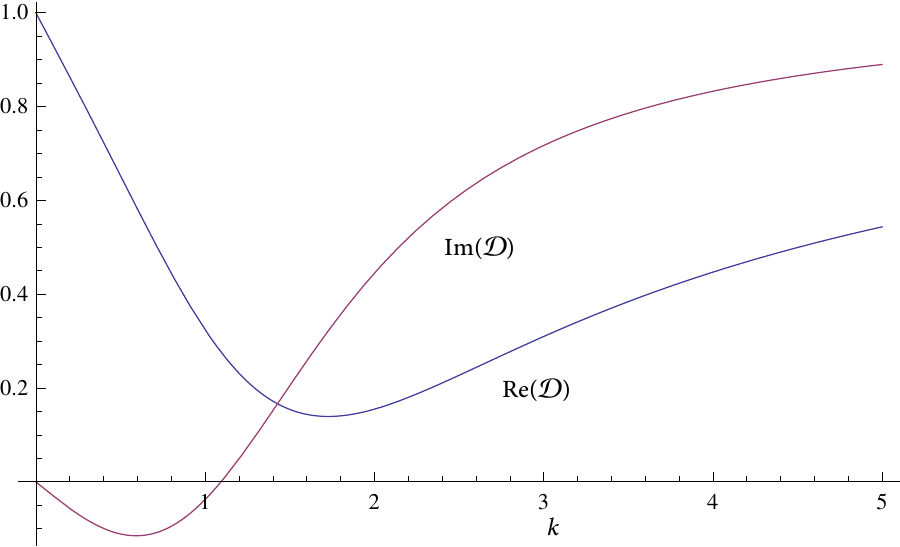}\ \ \ \
\includegraphics[width=7cm]{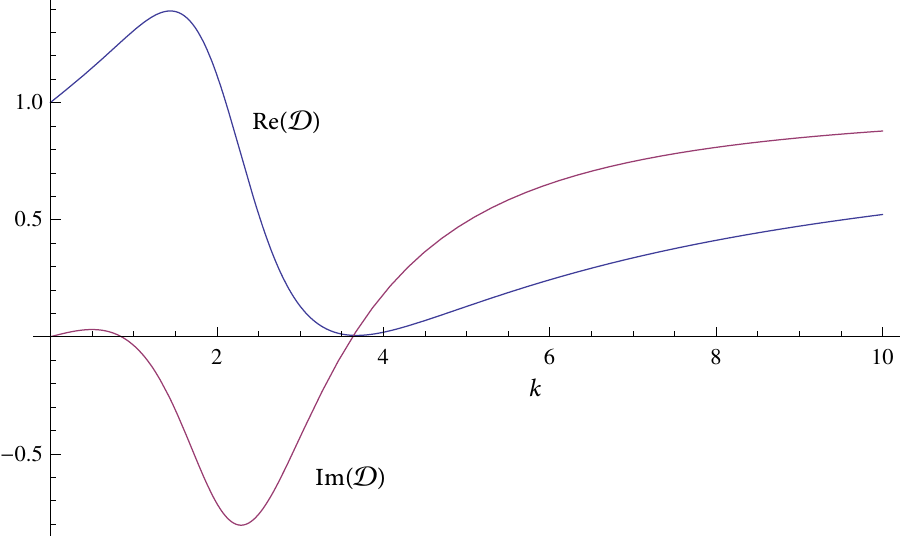}
\caption{These are plots of real (blue line) and imaginary (red line) part of ${\cal D}$ as functions of $k$ for some fixed $q$.  In the left figure, we have chosen $q=5$; in the right figure, we have $q=10$ and a Fermi surface emerges at $k=3.633657$. The black hole parameters are $r_+=2$ and $r_- =1/2$.}
\label{rmposnfsfs}
\end{figure}

   We first consider an example with both $r_\pm$ being positive.
To be specific, let us choose $r_+=2$ and $r_-=\fft12$. Thus the we have $\mu = \sqrt{q^2 - k^2}$.  For a given $q$, $k$ runs from 0 to $q$.  For small $q$, we find that there is no Fermi surface at all.  The left plot of Fig.~\ref{rmposnfsfs} shows both the real and imaginary parts of ${\cal D}$, as a function of $k$, for $q=5$.  It is clear that although the imaginary part of ${\cal D}$ passes through a zero, the real part is never particularly small.  The situation changes for larger $q$.  The right plot of Fig.~\ref{rmposnfsfs} has $q=10$, and an approximate Fermi surface emerges at $k=3.633657$.

\begin{figure}[ht]
\ \ \ \ \ \includegraphics[width=7cm]{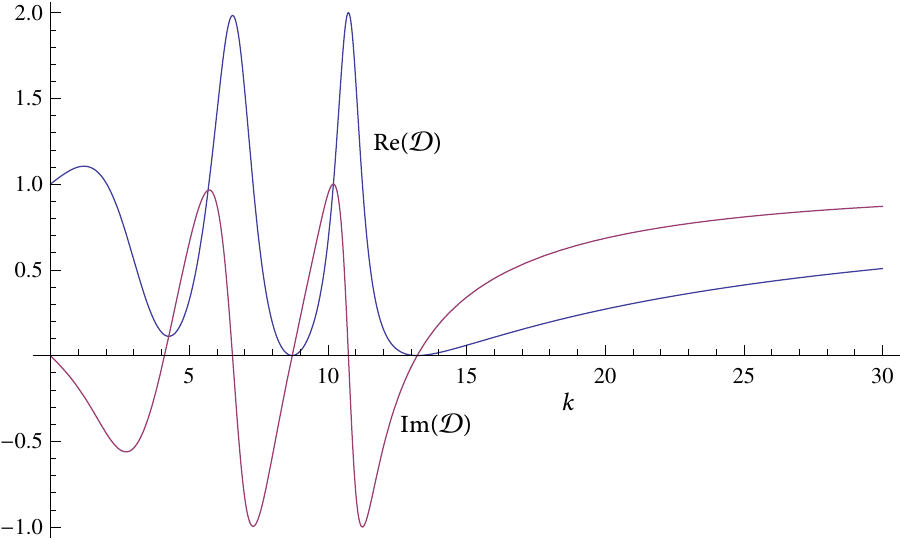}\ \ \ \
\includegraphics[width=7cm]{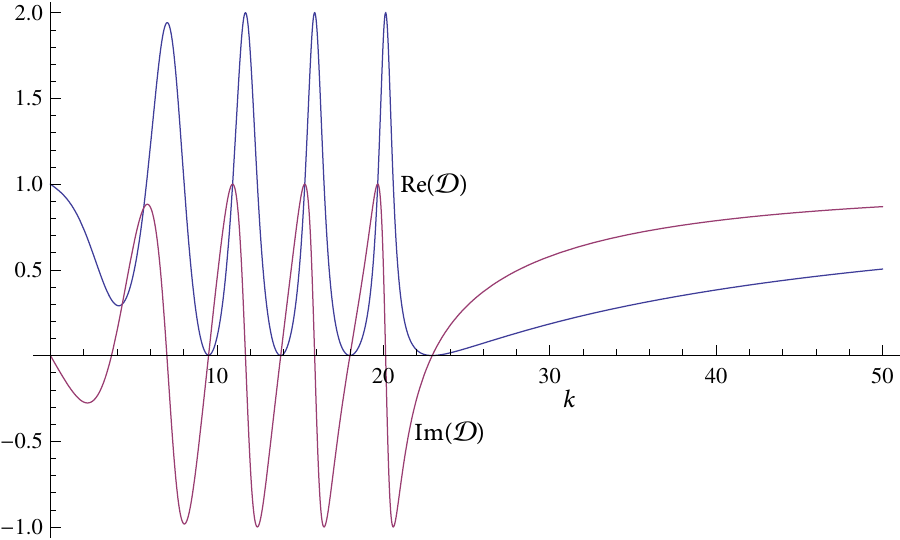}
\caption{These are plots of real (blue line) and imaginary (red line) part of ${\cal D}$ as functions of $k$ for some fixed $q$.  In the left plot, we have chosen $q=30$; in the right plot, we have $q=50$ and four Fermi surfaces occur at
$k_1=9.5042069087$, $k_2=13.8418437$, $k_3=17.998011219$, and $k_4=22.945804499291$. The black hole parameters are $r_+=2$ and $r_-=1/2.$}
\label{rmposmore}
\end{figure}

Let us increase the value of $q$, and we find that more and more Fermi surfaces emerge.  In Fig.~\ref{rmposmore}, we present the plots for $q=30$ and $50$.
\bigskip\bigskip

\bigskip

\begin{figure}[ht]
\center
\includegraphics[width=7cm]{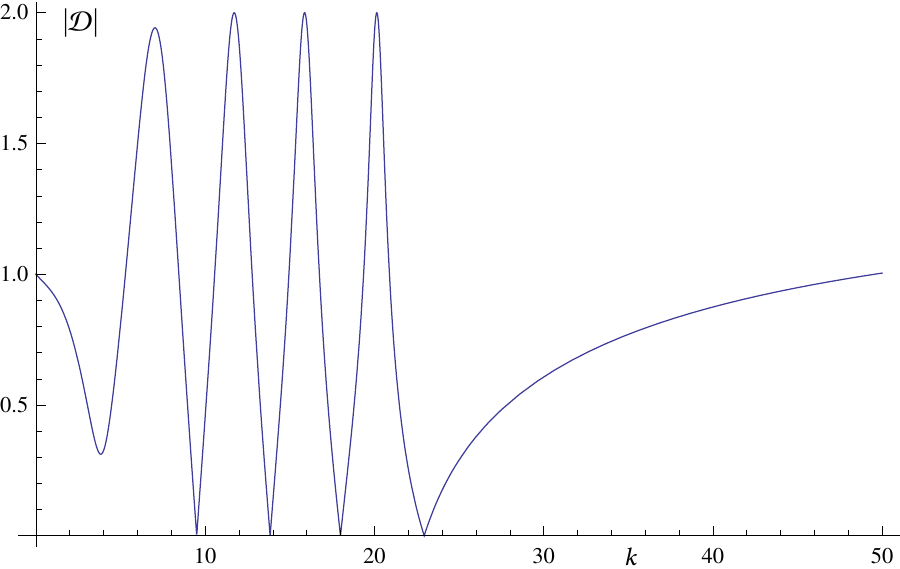}
\caption{This is a plot of $|{\cal D}|$ with respect to $k$, for $q=50$. The black hole parameters are $r_+=2$ and $r_-=1/2$.}
\label{rmposabs}
\end{figure}

It is worth remarking in the above graphs, when the imaginary (red) lines cross zeros, the real (blue) lines some times approaches a local minimum which is very small, but not necessarily zero literally.  For example, for $q=10$, the Fermi surface occurs at $k\sim 3.633657$ and we have ${\rm Re}({\cal D})\sim 0.0058$. For $q=50$, at $k_1$, we have ${\rm Re}({\cal D})\sim 0.0022$; at $k_2$, we have ${\rm Re}({\cal D})\sim 2.5\times 10^{-6}$; at $k_3$, we have ${\rm Re}({\cal D}) \sim 3\times 10^{-10}$; around at $k_4$, we have ${\cal D}\sim-2.2\times10^{-16} + 1.9\times 10^{-13} {\rm i}$.  Note that in this case, both real and imaginary parts cross zero around $k_4$.  Thus owing to the fact that ${\cal D}$ is complex, the $|{\cal D}|$ is not going to hit zero literally; however, the local minimums of $|{\cal D}|$ can be so small that these points should be viewed as Fermi surfaces.  In Fig.~\ref{rmposabs}, we plot the absolute value $|{\cal D}|$ for the $q=50$ example.

\begin{figure}[ht]
\ \ \ \ \ \includegraphics[width=7cm]{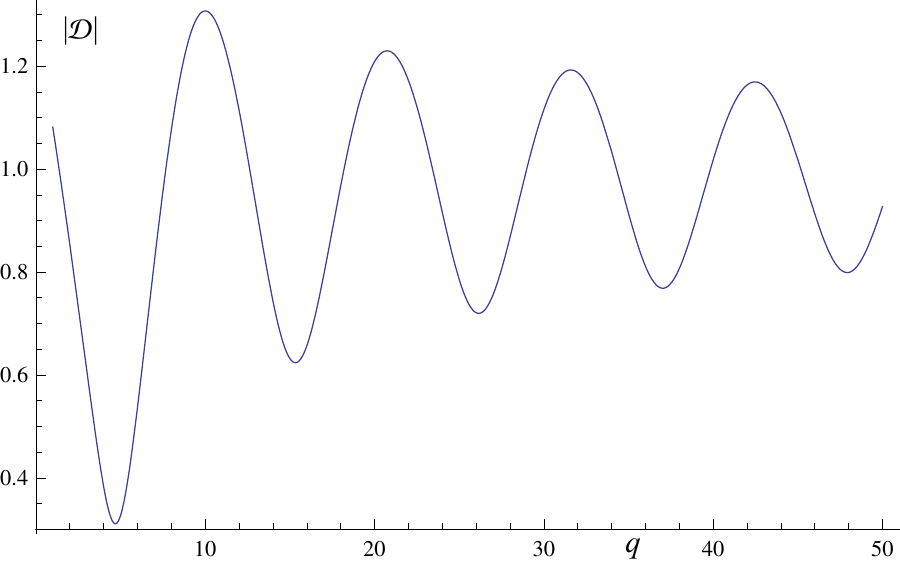}\ \ \ \
\includegraphics[width=7cm]{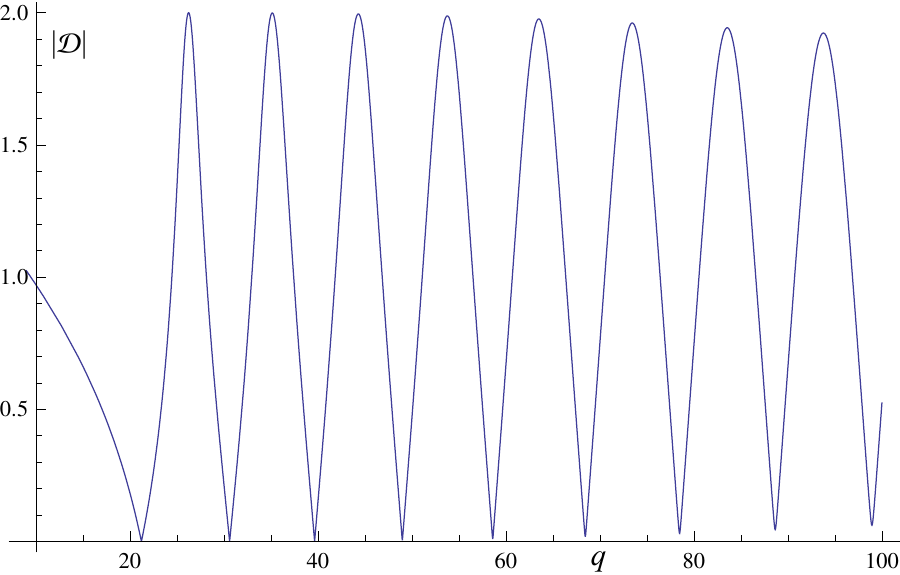}
\caption{These are plots of $|{\cal D}|$ with respect to $q$, for some fixed $k$. The left plot has $k=1$ and there is no Fermi surface.  In the right plot, we have $k=9$.  Multiple Fermi surfaces arise. The black hole parameters are $r_+=2$ and $r_-=1/2$.}
\label{rmposq}
\end{figure}

We have presented the effective Fermi surfaces so far for the given $q$'s.  It is instructive to present the pictures when $k$ is fixed and let $q$ run from $k$ to infinity.  In the left plot of Fig.~\ref{rmposq}, we have $k=1$, and we see that the minimal value of $|{\cal D}|$ is above 0.3 and hence there is no Fermi surface, even in the approximate sense.  In the right plot of Fig.~\ref{rmposq}, we let $k=9$, and we find that multiple but finite number of Fermi surfaces emerge.

\bigskip

\begin{figure}[ht]
\center
\ \ \ \ \ \includegraphics[width=6cm]{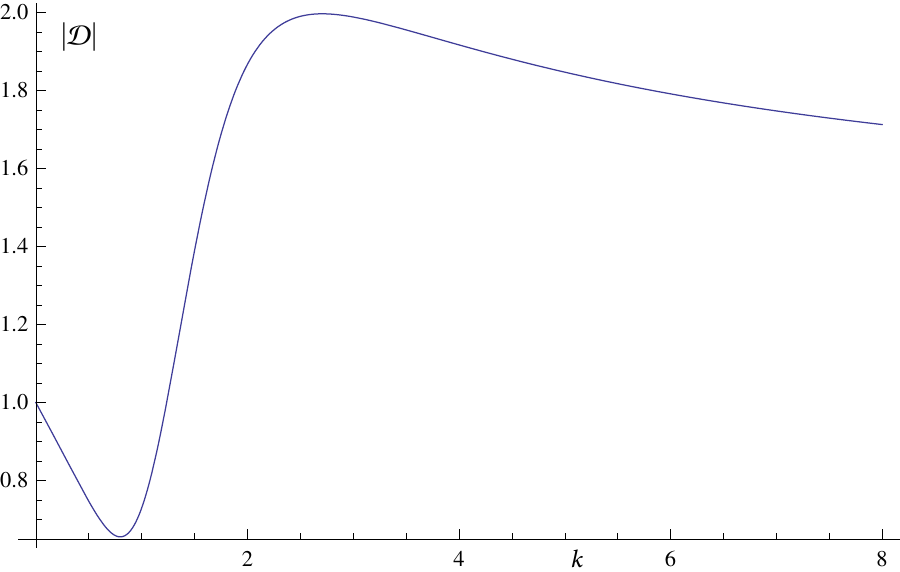}\ \ \ \
\includegraphics[width=6cm]{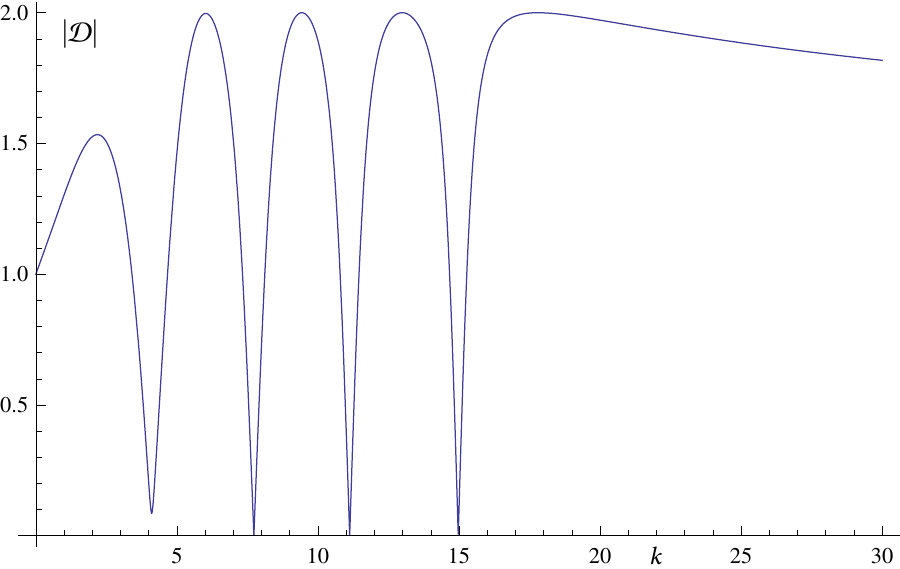}
\caption{These are two plots of $|{\cal D}|$ with respect to $k$, for some fixed $q$.  The left one has $q=4$ and the values of $|{\cal D}|$ lie roughly in the range of $(0.7,2)$.  The right one has $q=20$, and multiple Fermi surfaces indeed emerge. The black hole parameters are $r_+=1=-r_-$.}
\label{rmnegativefig}
\end{figure}

We now consider the second major case with $r_-<0$.  As a concrete example, we choose $r_+=1$ and $r_-=-1$, and hence we have $\nu={\rm i} \sqrt{k^2 + q^2}$.  Thus the parameters $k$ and $q$ can now run independently.  We find that for small $k$ or small $q$, there can be no Fermi surface.  In Fig.~\ref{rmnegativefig}, we give such an example with $q=4$ and the plot clearly shows the absence of a minimum that is close to zero.
When $k$ and $q$ are both sufficiently large, Fermi surfaces emerge, as can be seen in Fig.~\ref{rmnegativefig}.

\begin{figure}[ht]
\center
\includegraphics[width=7cm]{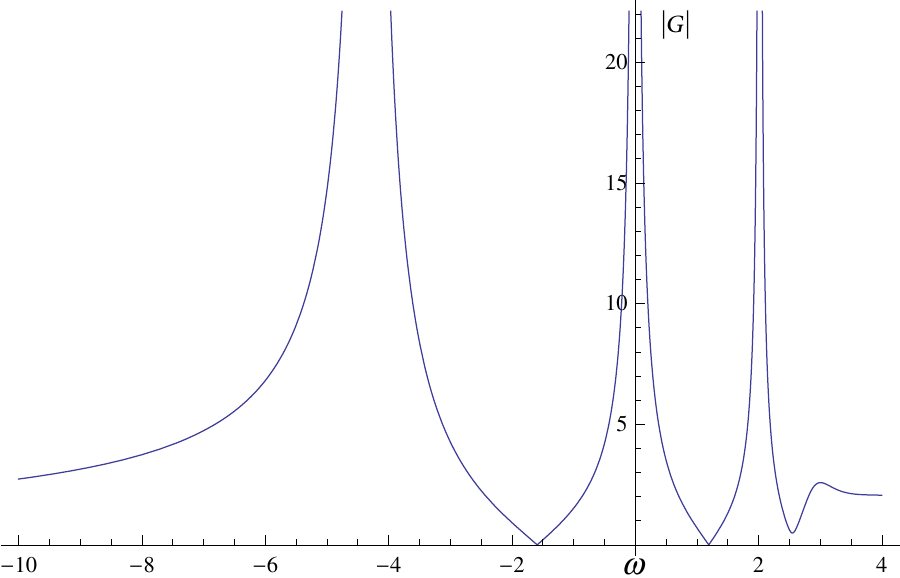}
\caption{This is the plot of $|G(\omega,k_F)|$ with respect to $\omega$ on a Fermi surface. The black hole is specified by $r_+=2$ and $r_-=1/2$.  The spinor charge is $q=50$.  The Fermi surface is at $k_F= 17.998011219$, for which the Green's function has a maximum around $10^9$.  Multiple and finite number of maxima arise as we change $\omega$. The maximum in the left is of $10^{10}$, the maximum in the middle with $\omega=0$ is of order $10^9$, whilst the maximum at the right is much smaller of 500.}
\label{omegafs}
\end{figure}

So far we have studied the Green's function with vanishing $\omega$, which allows us to determine the Fermi surfaces. It is also of great interest to study how the Green's function behaves when we change $\omega$.  Our analytic expression for general $(\omega, k)$ makes the task easy.  In Fig.~\ref{omegafs}, we present a graph on how the Green's function behaves under $\omega$ on a given Fermi surface $k=k_F$.  We see that multiple but finite maxima arise for $|G(\omega,k_F)|$.

\subsection{Near-extremal limit}

     As we have discussed in the previous subsections, there is only one
Fermi surface in the extreme limit, corresponding to $\nu=0$.  For a non-extremal black hole, multiple Fermi surfaces emerge for sufficiently large $q$.  It is of interest to study the near-extremal region.  Let us redefine the variables $r_+=r_0$ and $r_-=r_0 - \delta$.  The near extremal solution can be defined as having non-zero $\delta$ but with $\delta/r_0<<1$.

In the literatures, there have been discussions on obtaining decoupling limit of the near-extremal black holes.  For the black hole (\ref{genbulk0}) with (\ref{gensol1}), this can be achieved first by the following scalings and redefinitions:
\begin{equation}
r_+=r_0 + \epsilon \rho_0\,,\qquad r_-=r_0 - \epsilon \rho_-\,,\qquad r= r_0 + \epsilon \rho\,,\qquad t\rightarrow \fft{t}{\epsilon}\,.
\end{equation}
Then sending the parameter $\epsilon\rightarrow 0$ yields the following charged AdS$_2\times T^2$:
\begin{equation}
ds^2 = \fft{d\rho^2}{\rho^2-\rho_0^2} - (\rho^2 - \rho_0^2) dt^2 + r_0^2 (dx^2 + dy^2)\,,\qquad A=(\rho-\rho_0) dt\,.\label{ads2slice}
\end{equation}
The first two terms appear to describe a two dimensional non-extremal black hole; however, it is the AdS$_2$ metric.  In order to achieve this AdS$_2$ ``black hole'', the decoupling limit $\epsilon \rightarrow 0$ has to be taken so that the topology $\mathbb{R}^2\times T^2$ of the non-extremal black hole becomes AdS$_2\times T^2$. In this limit, we have $r_+-r_-= 2\epsilon \rho_0\rightarrow 0$, and hence the solution should really be viewed as the extremal limit, rather than the near-extremal solution.  In fact, in this topology-changing decoupling limit, all the thermodynamical quantities are identical to those of the extremal solution.  Thus the solution (\ref{ads2slice}) should be really viewed as a special slice of the AdS$_2\times T^2$ in the extremal limit.  In other words, as long as the solution is not extremal, no matter how near, the decoupling limit described above does not exist and the topology of the black hole remains $\mathbb{R}^2\times T^2$.

\begin{figure}[ht]
\ \ \ \ \ \includegraphics[width=7cm]{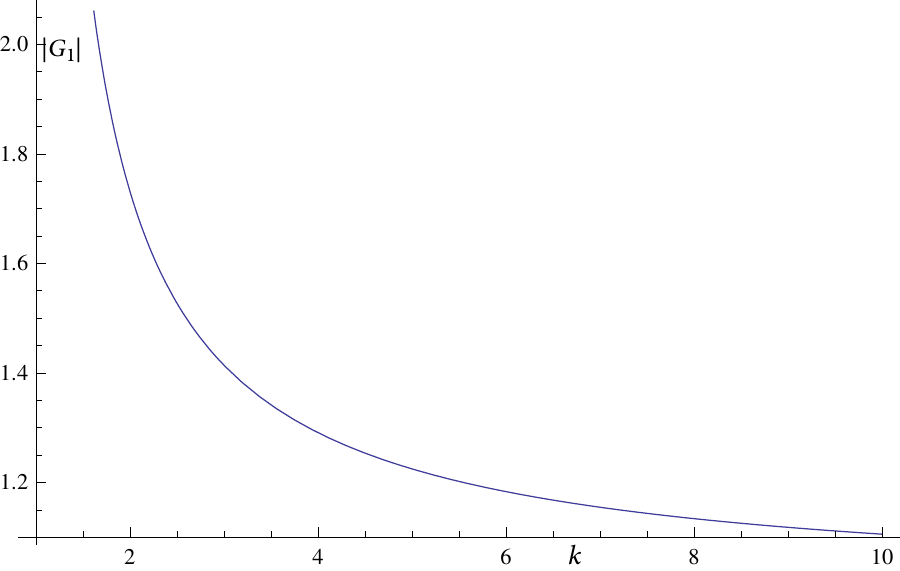}\ \ \ \
\includegraphics[width=7cm]{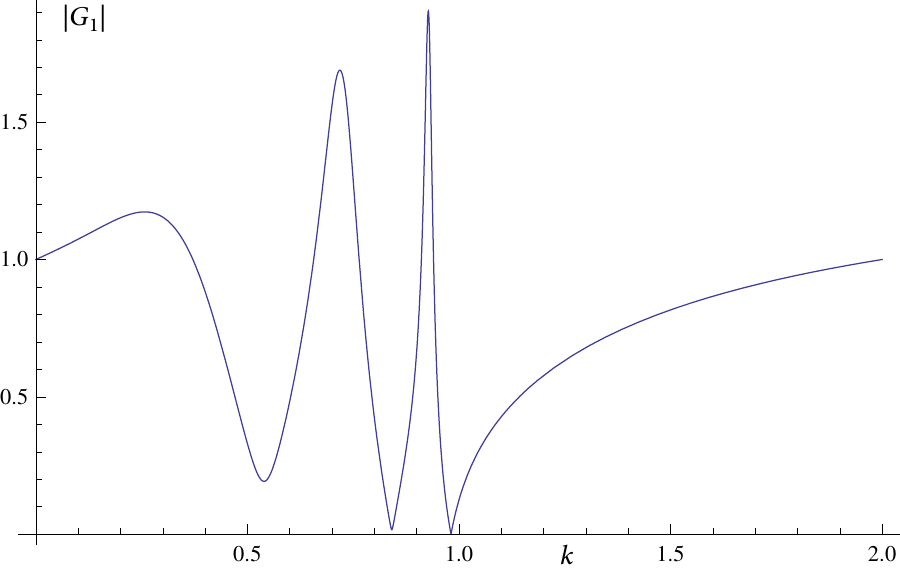}
\caption{In both figures, we have $\omega=0$, $r_+=1$, $r_-=1-10^{-4}$ and $q=1$. In the left figure, we plot $|G_1|$ vs.~$k$, with $k\ge k_0$.  In this case $\nu$ is real and consequently there is no Fermi surface. In the right figure, we plot $|G_1|$ vs.~$k$, with $0\le k\le k_0$. We have pure imaginary $\nu$ and Fermi surfaces emerge.}
\label{nefigs}
\end{figure}

   Now let us examine how the Green's function (\ref{bpsgreen1}) behaves in
the near-extremal region for the $\omega=0$ case.  Let us consider a concrete example, with $r_+=1$, $\delta=1/10^4$ and $q=1$.  For $\nu$ to be real, we must have $k\ge k_0=3\sqrt{1111}/100$.  We find that the absolute value $|G(k)|$ lies in between 16.0425 and 1 as $k$ runs from $k_0$ to $\infty$, suggesting that there is no Fermi surface for real $\nu$, as we see in the left figure in Fig.~\ref{nefigs}.  On the other hand, If we let $k$ be less than $k_0$ such that $\nu$ becomes pure imaginary, Fermi surfaces emerge, as we can see from the right figure in Fig.~\ref{nefigs}. Thus we see that the characteristics of the Fermi surface structure of near-extremal black holes is more or less the same as the generic non-extremal black holes.  This is consistent with the fact that the topology of the near-extremal solution is still a $\mathbb{R}^2\times T^2$, rather than the extremal AdS$_2\times T^2$.  It is worth pointing out that when $\delta$ is smaller so that the temperature is lower, the allowed the value of $q$ for the existence of a Fermi surface becomes smaller, as one would have expected.

\bigskip

Let us examine the Taylor expansion of the Green's function in terms of $\delta$.
Let $r_+=r_0$ and $r_-=r_0(1-\delta)$ with $\delta<<1$, and we keep the linear-order term in $\delta$.  Let us first consider $\nu$ is real.

\noindent{\bf Case 1:} $\nu > \fft12$ and $\nu\ne\,$half integers:

\begin{equation}
G=\sqrt{\fft{k+ q r_0}{k - q r_0}} + \fft{k q r_0\Big(r_0^2 - 8(k^2-q^2 r_0^2) - 2 r_0 \sqrt{k^2 - q^2 r_0^2}\Big)}{2(k-q r_0)^{\fft32}
\sqrt{k+ q r_0} \Big(4k^2 - (1 + 4q^2) r_0^2\Big)}\,\delta + {\cal O}(\delta^2)
\,.
\end{equation}
It is advantageous to express the result in terms of $\nu$ and $q$ variables,
given by
\begin{equation}
G=\fft{q \pm \sqrt{\nu^2 + q^2}}{\nu} + \fft{q r_0(\nu^2 + q^2 \pm q \sqrt{\nu^2 + q^2})}{\nu^2 (1 - 2\nu) r_0}\, \delta + {\cal O}(\delta^2)\,.
\end{equation}

\noindent{\bf Case 2:} $0<\nu < \fft12$:

\begin{eqnarray}
G &=& \sqrt{\fft{k+ q r_0}{k - q r_0}} + \fft{k}{k - q r_0}
\delta^{2\nu_0}\times \cr
&&\Big( {\rm i} +
\fft{2^{-4\nu_0} \Gamma(-2\nu_0) \Gamma (1 + 2\nu_0 - 2 {\rm i} q) \Gamma (2\nu_0 + 2 {\rm i} q) (\sin(2\pi q) - {\rm i} \sin (2\nu_0 \pi))}{2\pi \Gamma(2\nu_0)}\Big)\,,
\end{eqnarray}
where $\nu_0 = \sqrt{k^2 - q^2 r_0^2}/r_0$.

\begin{figure}[ht]
\center
\includegraphics[width=7cm]{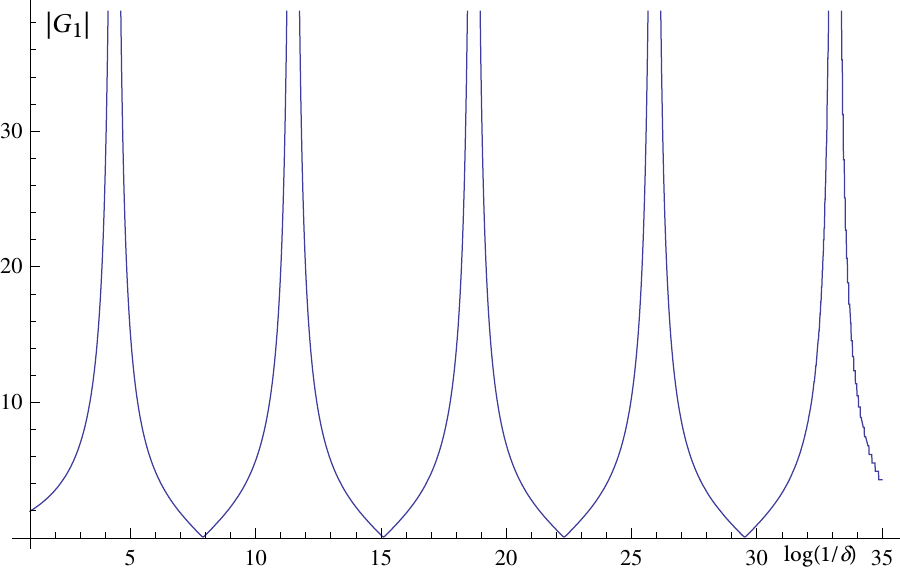}
\caption{Here we plot $|G_1|$ verse $\log(1/\delta)$.  It is clear that $|G_1|$ alternates between 0 and $\infty$ periodically in terms of $\log(1/\delta)$ as $\delta$ approaches zero. The parameters of the black hole are chosen to be $r_+=1$, $r_-=1 - \delta$. The Dirac fermion is chosen to have $q=1$ and $k=9/10$ so that $\nu$ is pure imaginary as $\delta$ becomes small.}
\label{infinitefs}
\end{figure}

If instead, we consider $\nu$ to be imaginary.  For a given $q$, no matter how small, as $\delta$ becomes smaller, there will be always a Fermi surface.
In fact there is an oscillatory factor $\delta^{\nu_0}$ in the Green's function, implies that as $\delta$ becomes smaller and smaller, more and more Fermi surfaces emerge for $k$ lying between $0$ and $q r_0$. The maximum value of $k$ for the Fermi surfaces becomes $k= q r_0$ in the extremal limit.  As $\delta$ approaches zero, the number of Fermi surfaces for $k< q r_0$ becomes infinite, signalling instability.  This is the physical origin of the earlier observation that the wave function near the horizon of the extremal black hole for imaginary $\nu$ has an infinite number of oscillations, as we saw in (\ref{infiniteosc}).  In Fig.~\ref{infinitefs}, we use an explicit example to plot this behaviour.

\subsection{Extracting the IR contributions}

Having obtained the Green's function for general $\omega$ and $k$, it is of interest to investigate the infrared (IR) and the ultra violet (UV) contributions.  For the extremal (or ``near-extremal'') solutions, the horizon geometry is AdS$_2\times T^2$, the IR contribution is governed by the boundary field theory associated with the AdS$_2$.  This makes it easier to extract the IR contribution. The situation is much subtler for the general non-extremal backgrounds.

   As demonstrated in \cite{flmv}, for extremal black holes with AdS$_2\times
T^2$ horizon, the Green's function for small $\omega$ can always be determined as long as $G(0,k)$ is known.  The wave functions determined by the matching procedure take the form \cite{flmv}
\begin{eqnarray}
&&\Psi=\eta_++\mathscr{G}(\omega)\eta_-\,,\label{generalform1}\\
&&\eta_{\pm}=\eta^{(0)}_{\pm}+\omega\eta^{(1)}_{\pm}+\omega^2\eta^{(2)}_{\pm}
+\cdots  \,,\label{generalform2}
\end{eqnarray}
where $\mathscr{G}(\omega)\sim\omega^{2\nu_k}$ is the Green's function in the IR AdS$_2$ region and the near-horizon asymptotic form of $\eta_{\pm}$
is given by nothing but the regular and singular solutions (\ref{extremalomega0}) at $\omega=0$, namely
\begin{eqnarray}
\eta^{(0)}_{\pm}\rightarrow v_{\pm}(r-r_0)^{\pm\nu_k}\,.
\end{eqnarray}
A procedure of determining $\eta_\pm$ order by order was developed in \cite{flmv}, which is applicable for a general class of extremal or near extremal black holes.

The wave function for our extremal black hole was given in (\ref{upmsol}).  We find that they can indeed be expressed in the form (\ref{generalform1}), but with $\eta_\pm$ given in close form in terms of $\omega$:
\begin{eqnarray}
u_1&=& \frac{  \Gamma\left(-2\nu_0\right)\Gamma\left(1-{{\rm i}{ q} }+\nu_0\right)}
{\Gamma\left(2\nu_0\right)\Gamma\left(1-{{\rm i}{ q} }-\nu_0\right)}
\left(-\frac{2{\rm i} \omega }{r_0}\right)^{2\nu_0}
\left(1-\frac{{r_0}}{r}\right)^{-\nu_0}e^{\frac{{\rm i} \omega }{r-r_0}}\times
\cr\!\!\!\!
&&\left[\fft{{\rm i} k}{r_0}{}_1F_1\left(1-{{\rm i}{ q} }+\nu_0;
1+2\nu_0;-z\right)
+\left(-{{\rm i}{ q} }-\nu_0\right)
{}_1F_1\left(-{{\rm i}{ q} }+\nu_0;
1+2\nu_0;-z\right)\right]
\cr\!\!\!\!&&\!\!\!\!
+
\left(1-\frac{{r_0}}{r}\right)^{{\nu_0}}e^{\frac{\ii \omega }{r-r_0}}\times
\cr
\!\!\!\! &&\left[\fft{{\rm i} k}{r_0}{}_1F_1\left(1-{\ii{ q} }-\nu_0;
1-2\nu_0;-z\right)
+\left(-{{\rm i}{ q} }+\nu_0\right)
{}_1F_1\left(-{{\rm i}{ q} }-\nu_0;
1-2\nu_0;-z\right)\right],\cr
u_2&=&
-{\rm i}
\frac{  \Gamma\left(-2\nu_0\right)\Gamma\left(1-{{\rm i}{ q} }+\nu_0\right)}
{\Gamma\left(2\nu_0\right)\Gamma\left(1-{\ii{ q} }-\nu_0\right)}
\left(-\frac{2{\rm i} \omega }{r_0}\right)^{2\nu_0}
\left(1-\frac{{r_0}}{r}\right)^{-{\nu_0}}e^{\frac{{\rm i} \omega }{r-r_0}}\times
\cr\!\!\!\!&&\left[\fft{{\rm i} k}{r_0}{}_1F_1\left(1-{\ii{ q} }+\nu_0;
1+2\nu_0;-z\right)
-\left(-{{\rm i}{ q} }-\nu_0\right)
{}_1F_1\left(-{{\rm i}{ q} }+\nu_0;
1+2\nu_0;-z\right)\right]
\cr\!\!\!\!&&\!\!\!\!
-\ii\left(1-\frac{{r_0}}{r}\right)^{{\nu_0}}e^{\frac{{\rm i} \omega }{r-r_0}}\times
\cr\!\!\!\!&&\left[\fft{{\rm i} k}{r_0}{}_1F_1\left(1-{{\rm i}{ q} }-\nu_0;
1-2\nu_0;-z\right)
-\left(-{{\rm i}{ q} }+\nu_0\right)
{}_1F_1\left(-{{\rm i}{ q} }-\nu_0;
1-2\nu_0;-z\right)\right]\,,
\end{eqnarray}
where $z=\frac{2{\rm i} \omega r}{r_0(r-r_0)}$ and $\nu_0=\sqrt{k^2/r_0^2 - q^2}$.  This provides a separation of the IR and UV contributions not only for small $\omega$, but also for general $\omega$ as well.

The situation is far subtler for the general non-extremal solutions, since the IR region now is not the AdS$_2\times T^2$, but $\mathbb{R}^2\times T^2$. Nevertheless, our analytic results for general $\omega$ and $k$ allow us to ``recreate'' the yet {\it unknown} perturbative approach and extract the IR contribution.  For simplicity, we present the recreation of the wave function for $u_-$ only.  The analogous result for $u_+$ follows straightforwardly.  We find the in-falling solution (\ref{rpminfalling}) can be split into two terms
\begin{eqnarray}
u_-&=&\fft{\Gamma(\fft12 - 2{\rm i} \Omega) \Gamma(-2\nu)}{\Gamma(-{\rm i} q - \nu)\Gamma(\fft12 + {\rm i} q +\nu - 2{\rm i}\Omega)} \left(\fft{r_+}{r_-} -1\right)^{-{\rm i} q + \nu} \left(1 - \fft{r_+}{r}\right)^{{\rm i}q -\nu -{\rm i} \Omega}\times\cr &&\qquad
\left(1 - \fft{r_-}{r}\right)^{-{\rm i}q + {\rm i}\Omega}
{}_2F_1(-{\rm i}q + \nu; \ft12 - {\rm i}q + \nu + {\rm i} \Omega; 1 + 2 \nu; z)\cr
&& + \fft{\Gamma(\fft12 - 2{\rm i} \Omega) \Gamma(2\nu)}{\Gamma(-{\rm i} q +\nu)\Gamma(\fft12 + {\rm i} q -\nu - 2{\rm i}\Omega)} \left(\fft{r_+}{r_-} -1\right)^{-{\rm i} q - \nu} \left(1 - \fft{r_+}{r}\right)^{{\rm i}q +\nu -{\rm i} \Omega}\times\cr &&\qquad
\left(1 - \fft{r_-}{r}\right)^{-{\rm i}q + {\rm i}\Omega}
{}_2F_1(-{\rm i}q - \nu; \ft12 - {\rm i}q - \nu + {\rm i} \Omega; 1 - 2 \nu; z)\,,
\end{eqnarray}
where $\Omega$ and $\nu$ are given by (\ref{omeganu}) and $z=r(r_--r_+)/(r_-(r-r_+))$.  It is then a straightforward exercise to obtain the extremal limit $r_+=r_0$, $r_-=r_0-\epsilon$ with $\epsilon\rightarrow 0$, and the two terms correspond to $\eta_\pm$ respectively.  Thus we expect that the Green's function in the IR region is given by the ratio of the above two terms, namely
\begin{equation}
{\cal G}(\omega)\sim \fft{\Gamma(-2\nu)\Gamma(-{\rm i} q + \nu)\Gamma(\fft12 + {\rm i}q -\nu - 2{\rm i} \Omega)}{\Gamma(2\nu)\Gamma(-{\rm i} q - \nu)\Gamma(\fft12 + {\rm i}q +\nu - 2{\rm i} \Omega)} \left(\fft{r_+}{r_-}-1\right)^{2\nu}\,.
\end{equation}
We find that this indeed gives rise to the same extremal result ${\cal G}(\omega)\sim \omega^{2\nu}$.  In fact,
we verify that this reproduces the Green's function of the IR region of the near-extremal limit of the RN black hole \cite{Faulkner:2011tm}.  It should be cautioned however that in non-extremal case, the $\omega$-dependence in the wave function appears in the parameters of the hypergeometric functions rather than in the variable $z$.  This makes it very difficult to extract the coefficients in the series expansion of $\omega$.  By contrast, such a dependence in the extremal case appears in the variable $z$ of the confluent hypergeometric function.  In obtaining the above result, we have assumed that there is no contribution to ${\cal G}(\omega)$ from the hypergeometric functions. If this conjecture is correct, we obtain the Green's function of the IR region in the general non-extremal case.  The result may also provide some clue as how to set the boundary condition for the $\omega=0$ wave solutions in the non-extremal background.

\section{Fermi surfaces from the general extremal black holes}

As we shall show in section 6, the wave solution for the most general black hole
is a Hern function whose properties are not conducive for discussing the Green's functions and Fermi surfaces in fine detail.  In this section we discuss the general extremal black hole (\ref{genextr}). The wave functions for general $(\omega,k)$ are still of complicated confluent Hern's functions; however, they reduce to hypergeometric functions when $\omega=0$.  This is enough to study the existence of Fermi surfaces.  We find that the general solution of $u_-$ is given by
\begin{eqnarray}
u_- &=& c_1 \Big(1-\fft{r_0}{r}\Big)^\nu \Big(1 - \fft{r_1}{r}\Big)^{{\rm i} q} {}_2F_1[\nu + {\rm i} q, \ft12 + \nu + {\rm i} q; 1 + 2\nu; -\ft{r_1(r-r_0)}{(r_0-r_1)r}]\cr
&&+c_2 \Big(1- \fft{r_0}{r}\Big)^{-\nu} \Big(1 - \fft{r_1}{r}\Big)^{{\rm i} q} {}_2F_1[-\nu + {\rm i} q, \ft12 -\nu + {\rm i} q; 1 - 2\nu; -\ft{r_1(r-r_0)}{(r_0-r_1)r}]\,.\label{genextrg}
\end{eqnarray}
where
\begin{equation}
\nu=\sqrt{\fft{k^2}{r_0(r_0-r_1)} - q^2}\,.\label{genextrnu}
\end{equation}
It is clear that for positive $\nu$, regularity of the wave function on the horizon $r=r_0$ requires that $c_2=0$. As we discussed in section 4, the ability to fix the boundary condition for $u_-$ on the horizon with $\omega=0$ is the property enjoyed only by the extremal solutions. If $\nu$ is imaginary, the wave function becomes oscillatory when $r\rightarrow r_0$ and we shall not consider such a case for the reasons that we have discussed in section 4.

    Having obtained $u_-$, we can easily obtain $u_+$ and hence the Green's
function, following the procedure spelled out in section 3.  We have
\begin{equation}
G_1=-G_2^{-1} ={\rm i} \fft{1 -{\rm i} \gamma}{1 + {\rm i} \gamma}\,,
\end{equation}
where
\begin{equation}
\gamma = \fft{(\nu+ {\rm i} q) r_0\, {}_2F_1[\fft12 + \nu + {\rm i} q, 1 + \nu + {\rm i q}; 1+ 2\nu; - \fft{r_1}{r_0-r_1}]}{k\, {}_2F_1[\nu + {\rm i}q, \ft12 + \nu + {\rm i} q; 1 + 2\nu, - \fft{r_1}{r_0-r}]}\,.
\end{equation}
This expression follows the same form of the Green's function in the earlier examples.  In fact, the Green's function with $\omega=0$ for the extremal solution is always real, provided that $\nu$ is real. This is because in the extremal black hole, the solutions for wave functions with real $\nu$ have two branches: one diverges on the horizon and the other converges. The boundary condition selects the converging branch for both $u_+$ and $u_-$, which has a consequence that they are complex conjugate to each other.  For a generic non-extremal solutions, both branches for $\omega=0$ wave solutions converge, and we have to use the $\omega\ne 0$ solutions to separate in-falling or outgoing branches and the in-falling solution should be chosen.  This has a consequence, as we have seen in the non-extremal example discussed in the previous section, that $u_\pm$ are not complex conjugate to each other and the Green's function $G(0,k)$ is hence not real.

The manifestly real expression for the Green's function (\ref{genextrg}) is given  by
\begin{equation}
G_1=-G_2^{-1} = \fft{{\rm Im} (Z)}{{\rm Re}(Z)}\,,
\end{equation}
where
\begin{equation}
Z=(1-{\rm i}) \Big(\nu- \fft{k}{\sqrt{r_0(r_0-r_1)}}+  {\rm i}q\Big)
\,{}_2F_1[\nu-{\rm i} q, \ft12 + \nu - {\rm i} q; 1 + 2\nu; - \fft{r_1}{r_0-r_1}]\,.
\end{equation}
It is easy to verify that when $q=0$, the Green's function becomes the identity, and hence there is no Fermi surface for $q=0$.  When $r_1=0$, we recover the Green's function given in (\ref{extrw0}), for which there is only one Fermi surface.

\begin{figure}[ht]
\center
\includegraphics[width=6cm]{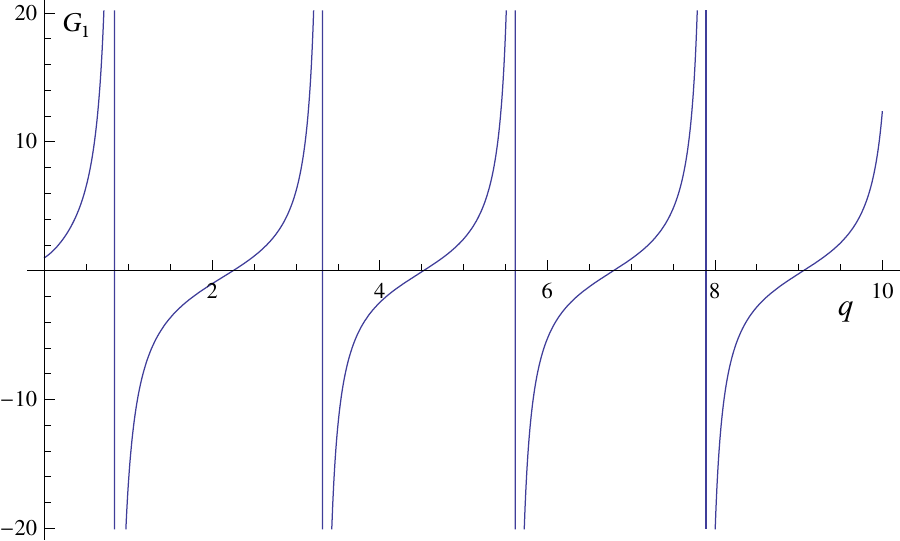}\ \ \ \ \ \ \
\includegraphics[width=6cm]{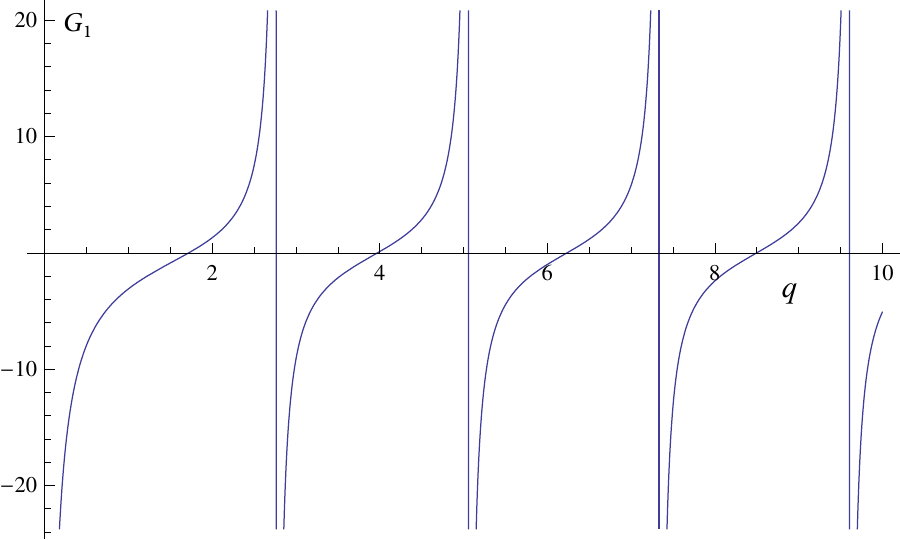}
\caption{For both graphs, we have the extremal black hole with $r_1=-1$ and $r_0=16/9$.  In the left figure, we fix $\nu=1/4$, so that $k=5\sqrt{1 + 16 q^2}/9$.  The plot describes $G_1$ with respect to $q$, and we see it oscillates between infinity and zero, corresponding to Fermi surfaces in standard and alternative quantisation respectively.  In the right plot, we consider the extreme non-Fermi liquids with $\nu=0$, corresponding to $k=20q/9$. Fermi surfaces emerge also.}
\label{extrfs}
\end{figure}

    It is of interest to note that when $\nu=1/4$, the function $Z$ becomes
simpler, giving
\begin{equation}
Z=(1 - {\rm i}) (1 - 4 {\rm i} q) (-1 - 4 {\rm i} q + \sqrt{1 + 16 q^2})
\sin\Big(\ft12 (1 + 4 {\rm i} q) \arctan \sqrt{\fft{r_1}{r_0-r_1}}\Big)\,,
\end{equation}
where we have dropped some inessential real multiplying factors and we have also chosen $k=\fft14 \sqrt{(1 + 16 q^2)r_0(r_0-r_1)}$ for solving $\nu=\fft14$.  Since $\nu=1/4$ corresponds to some Non-Fermi liquids, let us analyse the case further. For $r_1>0$, there appears to be no Fermi surface.  However, when $r_1<0$, Fermi surfaces emerge.  Let us consider a concrete example by choosing $r_1=-1$ and $r_0=16/9$.  For this choice of parameters, we have $k=5\sqrt{1 + 16 q^2}/9$. The Green's function is now given by
\begin{equation}
(G_1)^2=G_2^{-2} = \fft{1-\gamma}{4(1+\gamma)}\,,\qquad
\gamma=\fft{4q(4-5\cos(4q\log 2)) + 3 \sin(4q \log 2)}{\sqrt{16q^2 + 1} (5-\cos
(q \log 16))}\,.
\end{equation}
Note that $1-\gamma^2$ is a perfect square and the Green's function is manifestly real.  We present the plot of the Green's function as a function of $q$ in Fig.~\ref{extrfs}.  It oscillates between $\infty$ and zero, corresponding to Fermi surfaces in the standard and alternative quantization respectively.

Thus we see that for some appropriately chosen $q$ as indicated in Fig.~\ref{extrfs}, there exists a Fermi surface with $\nu=1/4$.  For either the standard or alternative quantization, $q$ must satisfy $1-\gamma^2=0$, implying
\begin{equation}
12 q \sin(4q \log 2) + 5 \cos(4q \log 2) - 4=0\,.
\end{equation}
Unfortunately, there appears to be no analytical solution to this equation, {\it albeit} it is rather a simple equation. The extreme case of the Non-Fermi liquids corresponds to $\nu=0$.  For the same black hole, this corresponds to $k=20q/9$.  We can see in Fig.~\ref{extrfs} that such extreme situation of Non-Fermi liquids can also arise in this extremal black hole.

\begin{figure}[ht]
\center
\includegraphics[width=5cm]{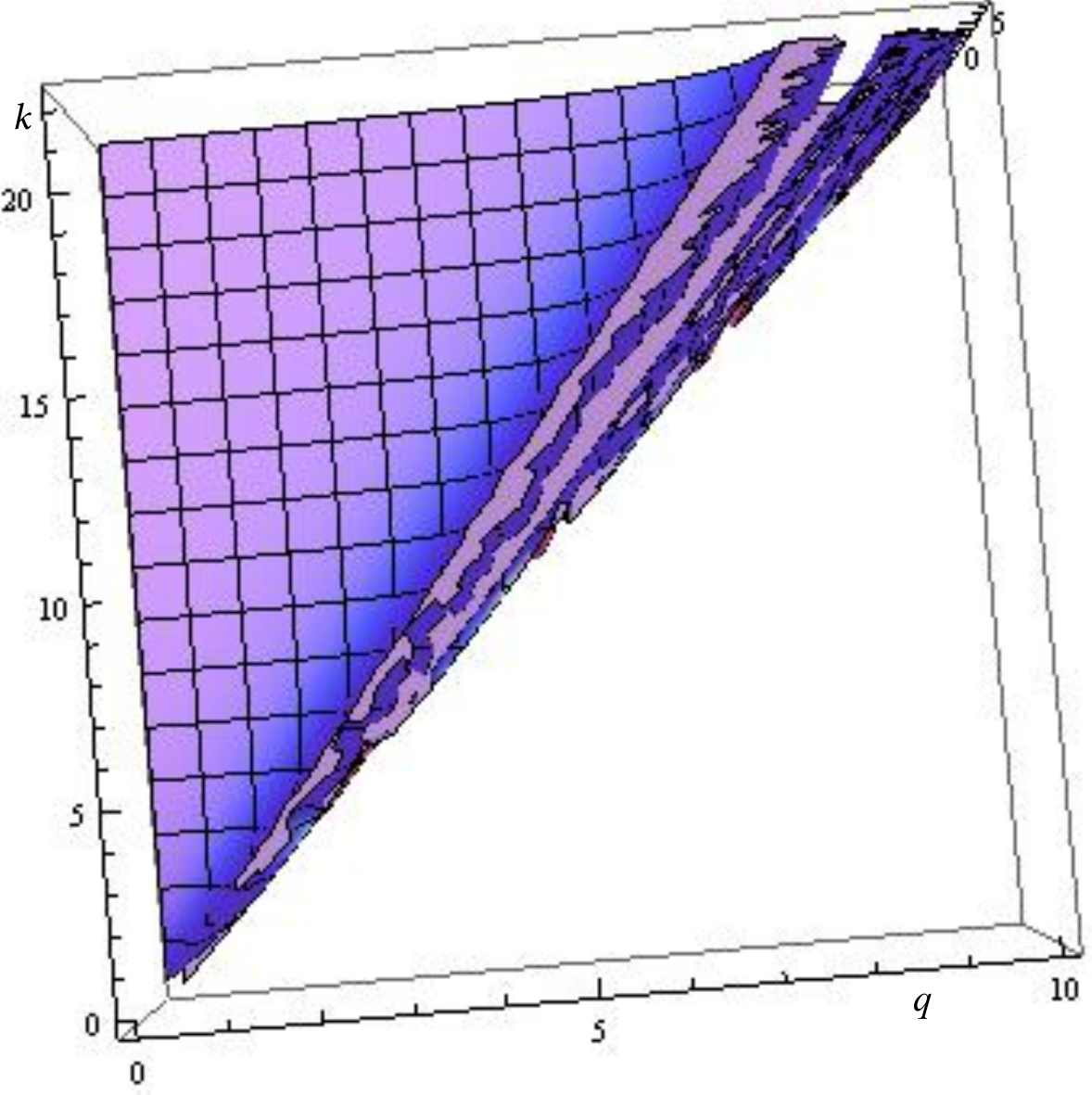}
\caption{The black hole parameters are $r_1=-1$ and $r_0=16/9$. This is the 3D plot of $G$ in the $(k,q)$-plane, viewed from the top. ($k$ runs from 0 to 20 and $q$ runs from 0 to 10.) Interestingly the Fermi surfaces seem to lie naturally in the linear line of $k$ of $q$, with the diagonal line corresponding to $\nu=0$.}
\label{extr3d}
\end{figure}

In the above we discussed a special case with $\nu=\fft14$, corresponding to some non-Fermi liquids.  We also saw that the extreme non-Fermi liquids with $\nu=0$ can also arise.  By numerating examples, we find that it appears that Fermi surfaces with all possible $\nu$ can arise in the $(k,q)$ plane.  For the black hole with $r_1=-1$ and $r_0=16/9$, we present the 3D plot in Fig.~\ref{extr3d}.  Comparing to
the similar plot in the non-extremal case, we find that the Green's function in the extremal background is much simpler than that in the non-extremal background.

The situation for the general case is similar.  We find no evidence for Fermi surfaces when $r_1$ is positive. However, when $r_1$ is none positive, Fermi surfaces emerge.  When $r_1=0$, it reduces to the extremal solution discussed in section 4. In this case, there is only one Fermi surface, with $k=q r_0$ for the standard quantization and $k=-q r_0$ for the alternative quantization.  When $(-r_1)/r_0<<1$, there is still only one Fermi surface in each quantization, with $k\sim\pm q (r_0 - \fft12 r_1)$.  The more negative $r_1$ is, the more number of Fermi surfaces we can have for each given $q$.

\begin{figure}[ht]
\center
\includegraphics[width=7cm]{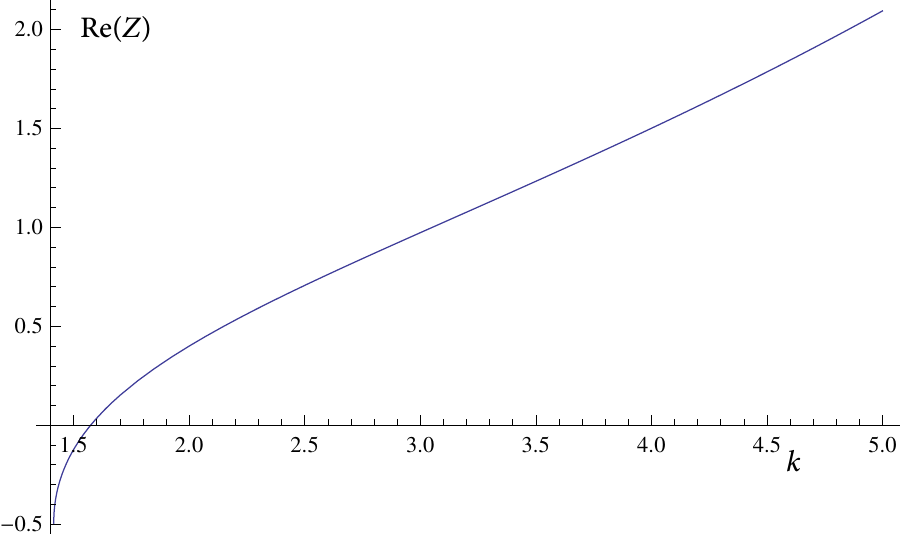}\ \ \ \
\includegraphics[width=7cm]{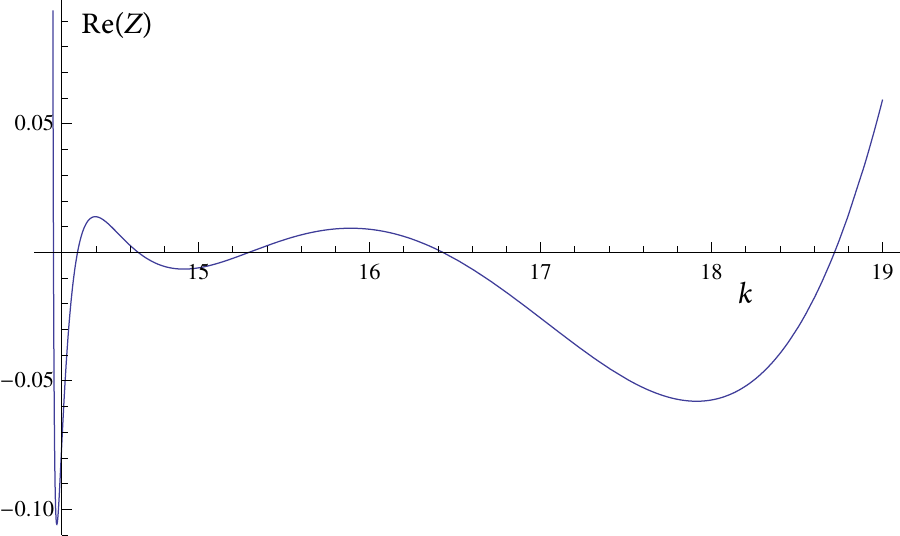}
\caption{In both graphs, we have $r_0=1$ and $r_1=-1$. The plots are ${\rm Re}(Z)$ with respect to $k$ whose minimum value is $\sqrt{2} q$ for real $\nu$.  The zeros of the plots indicate the Fermi surfaces. The left figure corresponds to $q=1$, which only gives rise to a single Fermi surface.  The right figure corresponds to $q=10$, and six Fermi surfaces emerge.  For this extremal black hole, we find that the minimum Fermi surface has $\nu<\fft12$.}
\label{extrf23}
\end{figure}

      For a given black hole with $r_0$ and a negative $r_1$, the existence of
Fermi surfaces favors a larger $q$.  We find that there is always at least one Fermi surface and the number of the Fermi surfaces increases when $q$ becomes larger.  We demonstrate this with a concrete example.  Let us consider $r_0=1$ and $r_0=-1$.  The Fig.~\ref{extrf23} shows two plots of ${\rm Re}(Z)$ with respect to $k$ which runs from $\sqrt{2} q$ to $\infty$.  It is clear that for smaller $q=1$ in the left figure, there is only one Fermi surface and for larger $q=10$, there are six Fermi surfaces arising.

As we mentioned earlier, when $-r_1/r_0\rightarrow  0^+$, there is only one Fermi surface with $\nu\rightarrow 0$.  If we increase the value of $-r_1/r_0$, more Fermi surface will arise.  For fixed $q$, the minimum $\nu$ increases and can eventually pass $\nu=\fft12$.  However, for such black holes, we can always lower the $q$ value so that the minimal $\nu$ is less than $\fft12$.

\section{Green's function from the most general black holes}

In this section, we study the Dirac equation (\ref{diracE0}) on the most general black hole (\ref{rieg}) with $T^2$ topology ($\varepsilon=0$).  It is advantageous to use the three roots $r_1,r_2$ and $r_3\equiv r_0$ of $f$ as the parameters of the solution.  In this parametrisation, the solution is given by
\begin{eqnarray}
&&f=\ft1{r} (r-r_1)(r-r_2)(r-r_0)\,,\qquad A= Q\,(\ft{1}{r_0}-\ft{1}{r})\, dt\,,\cr
&&Q^2=(r_1 r_2)^2 + (r_0r_1)^2 + (r_0r_2)^2 - r_1 r_2 r_0(r_1 + r_2 + r_0)\,.\label{mostgensol}
\end{eqnarray}
Note that without loss of generality, we have chosen $r_0>0$ to represent the horizon radius.  If both $r_1$ and $r_2$ are complex numbers, they must be conjugate to each other; if they are real, neither of them is bigger than $r_0$.  Following the procedure in section 3, we find that
\begin{equation}
\lambda_1(r) = {\rm i} \fft{(\omega + q Q/r_0) r - q Q}{(r-r_1)(r-r_2)(r-r_0)}\,,\qquad \lambda_2(r) =- \fft{{\rm i} k}{r\sqrt{r(r-r_1)(r-r_2)(r-r_0)}}\,.
\end{equation}
Note that $q Q$ appears in the Dirac equation together, implying that the complicated expression for $Q$ in (\ref{mostgensol}) can be absorbed in $q$ such that $qQ$ remains a simple expression. Let us first solve for $u_-$.  In order to simplify the equation, we let
\begin{eqnarray}
u_-=(r-r_1)^{-{\rm i} (\widetilde Q+\Omega_1)} (r-r_2)^{{\rm i} (\widetilde Q- \Omega_2)}
(r-r_0)^{-{\rm i} \Omega_0} v(z)\,, \label{Heunmani}
\end{eqnarray}
where
\begin{eqnarray}
&&z=\fft{r_1(r-r_0)}{r(r_0-r_1)}\,,\qquad \widetilde Q= \fft{q Q}{r_0(r_1-r_2)}\,,\qquad \Omega_1=\fft{\omega r_1}{(r_1-r_0) (r_1-r_2)}\,,\cr
&& \Omega_2=\fft{\omega r_2}{(r_2-r_0) (r_2-r_1)}\,,\quad
\Omega_0=\fft{\omega r_0}{(r_0-r_1) (r_0-r_2)}=\fft{\omega}{4\pi T}\,.
\end{eqnarray}
Here $T$ is the temperature of the black hole, whose thermodynamics was analysed in section 2. Note that we have the identity $\Omega_1 + \Omega_2 + \Omega_0=0$.  Thus the three factors $(r-r_i)$ with ($i=0,1,2)$ in (\ref{Heunmani}) can be simultaneously replaced by $(1-r_i/r)$.
Substituting (\ref{Heunmani}) into the wave equation of $u_-$, we find that $v(z)$ satisfies the following Heun's equation
\begin{eqnarray}
v''(z) +
\left[\ft{\fft12 - 2 {\rm i} \Omega_0}{z}+ \ft{\fft12 - 2{\rm i} (\widetilde Q + \Omega_1)}{z-1}
+ \ft{\fft12 + 2{\rm i} (\widetilde Q -\Omega_2)}{z-a} \right]
v'(z) - \frac {b} {z(z-1)(z-a)} v(z) = 0\,,
\end{eqnarray}
with
\begin{eqnarray}
&&a=\fft{(r_0-r_2)r_1}{(r_0-r_1)r_2}\,,\qquad  b=-\fft{k^2}{(r_0-r_1) r_2}\,.
\end{eqnarray}
Therefore, the wave function $u_-$ can be solved in terms of the following general Heun's function $H\!\ell$:
\begin{eqnarray}
u_-&=& c_1 (r-r_1)^{-{\rm i} (\widetilde Q+\Omega_1)} (r-r_2)^{{\rm i} (\widetilde Q- \Omega_2)}
(r-r_0)^{-{\rm i} \Omega_0} \times \cr
&& H\!\ell\Big( a,b ; 0, \ft12, \ft12 - 2 {\rm i} \Omega_0, \ft12 - 2{\rm i} (\widetilde Q + \Omega_1); -z\Big)\cr
&& + c_2 r^{-\fft12 -2{\rm i}\Omega_0} (r-r_1)^{-{\rm i} (\widetilde Q+\Omega_1)} (r-r_2)^{{\rm i} (\widetilde Q- \Omega_2)}
(r-r_0)^{\fft12+{\rm i} \Omega_0} \times\cr
&& H\!\ell\Big( a, \beta; \ft12 + 2 {\rm i} \Omega_0, 1+ 2 {\rm i} \Omega_0, \ft32+ 2 {\rm i} \Omega_0, \ft12 - 2{\rm i} (\widetilde Q + \Omega_1); -z\Big)\,,\label{genwavegenbh}
\end{eqnarray}
where
\begin{equation}
\beta=b - (\ft12 + 2{\rm i} \Omega_0)\Big(\fft{2{\rm i} q Q}{r_2(r_0-r_1)} + \fft{2r_1 r_2 - (r_1+r_2) r_0}{2 r_2 (r_0-r_1)}(1 + 4{\rm i}\Omega_0)\Big)\,.
\label{beta}
\end{equation}
As we have mentioned in section 4.4, the parameter combination $\nu$, which plays an important role in extremal solutions, has no apparent physical significance in the non-extremal backgrounds.  In fact it does not appear in the above wave solutions at all.  Comparing to (\ref{genextrnu}) for the general extremal case corresponding to having $r_2= r_0$, we expect that the $\nu$ may arise in the following combination
\begin{equation}
\nu^2\Big|_{\rightarrow\rm extremal} = -(b + \widetilde Q^2)\Big|_{r_2\rightarrow r_0}\,.
\end{equation}
Thus we see clearly that in the non-extremal background, such a combination does not appear in the solution and hence it should play no particular role.  There is no physical reason to impose the reality condition. It follows that there is no constraint on $(k,q)$ in the non-extremal case.  The emergence of the $\nu$-like quantity in the subclass of non-extremal black holes discussed in section 4 was pure accidental.

On the horizon $r=r_0$, corresponding to $z=0$, the Heun's functions in (\ref{genwavegenbh}) all reduce to identity, and hence the $c_1$ term corresponds to the in-falling mode and the $c_2$ term is the outgoing mode.  For the black hole background, it is necessary to choose the in-falling solution only, and hence we set $c_2=0$. Having obtained the in-falling solution for $u_-$, the function $u_+$ can then be read off from (\ref{upfromum}). (It actually requires some derivative identity given presently.) Thus the full set of in-falling solutions are
\begin{eqnarray}
u_-&=& (1-\ft{r_1}{r})^{-{\rm i} (\widetilde Q+\Omega_1)}\, (1-\ft{r_2}{r})^{{\rm i} (\widetilde Q- \Omega_2)}\,
(1-\ft{r_0}{r})^{-{\rm i} \Omega_0} \times \cr
&& H\!\ell\Big( a,b ; 0, \ft12, \ft12 - 2 {\rm i} \Omega_0, \ft12 - 2{\rm i} (\widetilde Q + \Omega_1); -z\Big)\,,\cr
u_+&=& c\,(r-\ft{r_1}{r})^{{\rm i} (\widetilde Q+\Omega_1)}\, (1-\ft{r_2}{r})^{-{\rm i} (\widetilde Q- \Omega_2)}\,
(1-\ft{r_0}{r})^{\fft12-{\rm i} \Omega_0} \times\cr
&& H\!\ell\Big( a, \beta^*; \ft12 - 2 {\rm i} \Omega_0, 1- 2 {\rm i} \Omega_0, \ft32- 2 {\rm i} \Omega_0, \ft12 + 2{\rm i} (\widetilde Q + \Omega_1); -z\Big)\,,
\end{eqnarray}
where $\beta^*$ is the complex conjugate of $\beta$ given in (\ref{beta}) and $c$ is a complex constant:
\begin{equation}
c=\fft{\rm i}{k} (1-\ft{r_1}{r_0})^{\fft12 -2{\rm i} (\widetilde Q + \Omega_1)}\, (1 - \ft{r_2}{r_0})^{\fft12 + 2{\rm i} (\widetilde Q - \Omega_2)}.
\end{equation}
This constant coefficient can only be obtained from evaluating (\ref{upfromum}), even though the structure of the solution for $u_+$ can be obtained right away by the fact that $u_+$ takes the complex conjugate form of $u_-$.  It follows from (\ref{gfformula}) that the Green's function can now be obtained straightforwardly:
\begin{equation}
G(\omega, k) = {\rm i} \fft{1 - \gamma(\omega,k)}{1 + \gamma(\omega,k)}\,,\label{mostgengreen}
\end{equation}
where
\begin{equation}
\gamma(\omega,k) = \fft{c\, H\!\ell\Big( a, \beta^*; \ft12 - 2 {\rm i} \Omega_0, 1- 2 {\rm i} \Omega_0, \ft32- 2 {\rm i} \Omega_0, \ft12 + 2{\rm i} (\widetilde Q + \Omega_1); -\fft{r_1}{r_0-r_1}\Big)}{H\!\ell\Big( a,b ; 0, \ft12, \ft12 - 2 {\rm i} \Omega_0, \ft12 - 2{\rm i} (\widetilde Q + \Omega_1); -\fft{r_1}{r_0-r_1}\Big)}\,.
\end{equation}
Compared to the hypergeometric functions, Heun's functions are much less studied.  In particular, in the Mathematica package, the Heun's functions are not yet coded.  In the Maple package, although the function is built in, very limited properties are coded.  Note that when $r_1$ or $r_2$ vanishes, the Heun's functions are reduced to hypergeometric functions and we obtain the results for the ``BPS'' black holes, discussed in section 4.  If we let $r_1$ or $r_2$ to be $r_0$, we obtain the results for the most general extremal solution. The limit is rather subtle since several arguments in the Heun's function diverge. We find that the Heun's functions become some confluent Heun's functions $H\!\ell c$ in this limit.  The proper procedure requires the identity (\ref{nonidentity}) with precise coefficients which are yet unknown in the literature.  We present the general solution for $u_-$ for general extremal solution for which $r_2=r_0$:
\begin{eqnarray}
u_-\!\!\!&=&\!\!\!c_2 (1-\ft{r_1}{r})^{ -\fft12 + {\rm i} (\widehat Q - \widehat\Omega)} (1-\ft{r_0}{r})^{-{\rm i} (\widehat Q - \widehat \Omega)} H\!\ell c \Big(-2{\rm i} \widehat \Omega, -\ft12, -\ft12 + 2{\rm i}(\widehat Q -\widehat \Omega),\delta,\eta;-\hat z\Big)\cr
\!\!\!&&\!\!\! + c_1 (1-\ft{r_1}{r})^{ {\rm i} (\widehat Q - \widehat\Omega)} (1-\ft{r_0}{r})^{-{\rm i} (\widehat Q - \widehat \Omega)} H\!\ell c \Big(-2{\rm i} \widehat \Omega, -\ft12, -\ft12 + 2{\rm i}(\widehat Q -\widehat \Omega),\delta,\eta;-\hat z\Big),\label{genextrsol}
\end{eqnarray}
where
\begin{eqnarray}
&&\widehat Q=\fft{qQ}{r_0(r_0-r_1)}\,,\qquad
\widehat \Omega=\fft{\omega r_1}{(r_0-r_1)^2}\,,\qquad z=\fft{(r_0-r_1)r}{r_1(r-r_0)}\,,\cr
&&\delta={\rm i}\widehat \Omega (1 + 2{\rm i} (\widehat Q - \widehat \Omega)\,,\qquad
\eta= \ft38 - \ft12{\rm i} \widehat Q - \fft{k^2}{r_0(r_0-r_1)}\,,
\end{eqnarray}
In order to proceed further, it is necessary to fix the horizon condition on this general wave solution for the two-parameter extremal black hole.  However, on the horizon, we have $z\rightarrow \infty$, it is thus necessary to change the variable $z$ to $1/z$.  Unfortunately, we have not found in the literature the necessary identities to perform this transformation.  We leave this problem open for future investigation.

\begin{figure}[ht]
\center
\includegraphics[width=6cm]{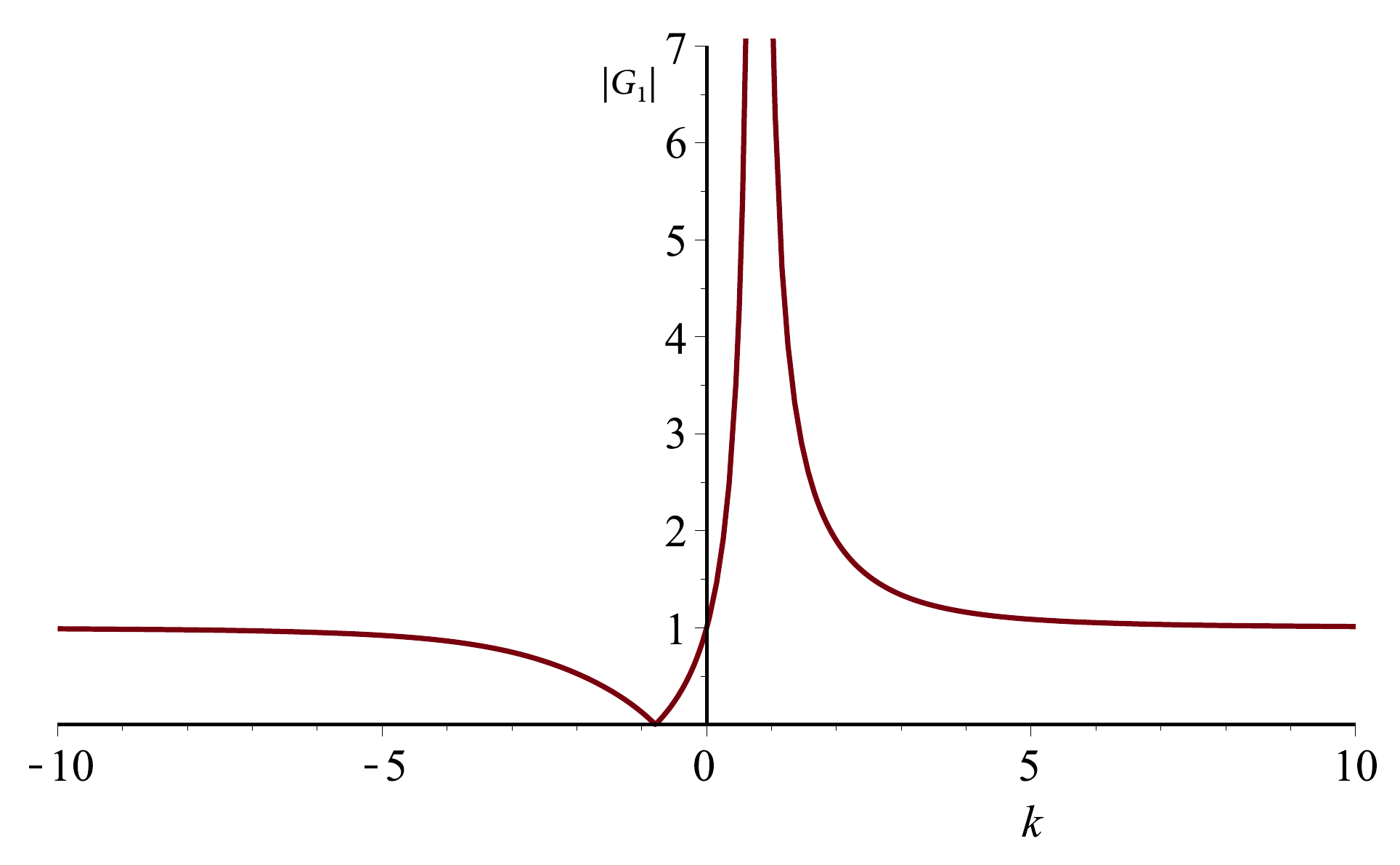}\ \ \ \
\includegraphics[width=6cm]{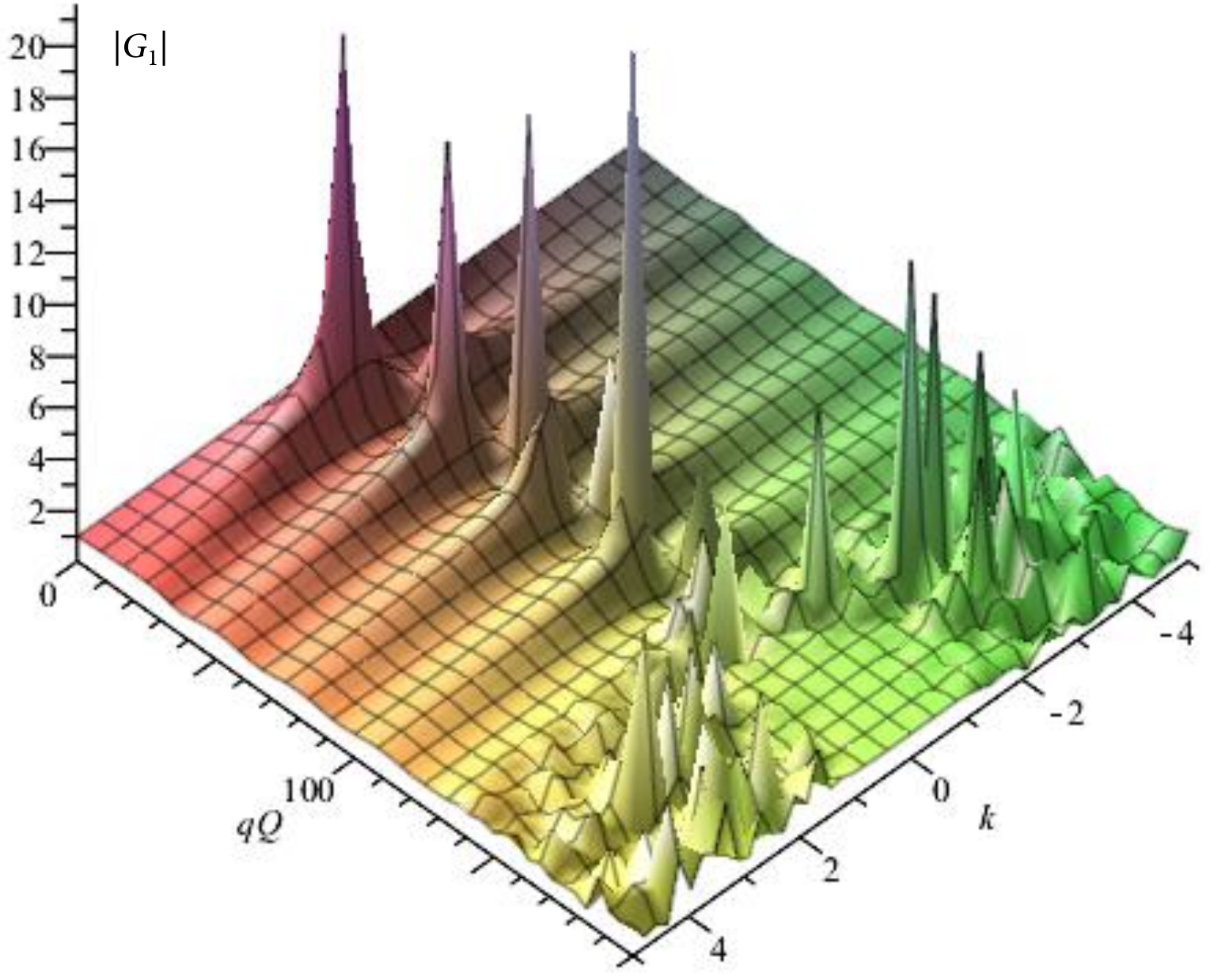}
\caption{For both plots, the black hole parameters are $r_0=3$, $r_1=1$ and $r_2=-2$.  The left is the plot of $|G_1|$ with respect to $k$, with $qQ=11.75535749$. The maximum occurs at $k_F=0.7903492261$ and it is of order $10^9$.  The minimum is of order $10^{-8}$.  The right is the three-dimensional plot of $|G_1|$ with both $k$ and $qQ$ variables. Many spiked maxima arise.}
\label{fsinmostgen}
\end{figure}

In order to determine the Fermi surfaces from (\ref{mostgengreen}), we set $\omega=0$, and the Green's function becomes
\begin{eqnarray}
G(0, k) &=& {\rm i} \fft{1 - \gamma(k)}{1 + \gamma(k)}\,,\cr
\gamma(k) &=&  \fft{{\rm i} (r_0-r_1)^{\fft12 -2{\rm i} \widetilde Q}\, (r_0 - r_2)^{\fft12 + 2{\rm i} \widetilde Q}\, H\!\ell\Big( a, \beta^*; \fft12, 1, \fft32, \fft12 + 2{\rm i} \widetilde Q; -\fft{r_1}{r_0-r_1}\Big)}{k r_0\,H\!\ell\Big( a, b ; 0, \fft12, \fft12, \fft12 - 2{\rm i}\widetilde Q; -\fft{r_1}{r_0-r_1}\Big)}\,,
\end{eqnarray}
where $\beta$ is given in (\ref{beta}) with $\omega=0$.  Analogous to the simpler non-extremal example discussed in section 4, $G(0,k)$ is not real and there is no literal divergence .  However, there are effective divergencies corresponding large spiked maxima of $|G(0,k)|$. To demonstrate that the above Green's function can indeed produce effective Fermi surfaces, let us consider a concrete example.  Let us consider $r_0=3$, $r_1=1$ and $r_2=-2$.  A fermi surface occurs at $qQ=11.75535749$ and $k_F=0.7903492261$.  The absolute value of the Green's function at this point is of order $10^9$.  We present some graphs of $|G(0,k)|$ in Fig.~\ref{fsinmostgen}. It is easy to see that the maxima associated with the effective Fermi surface is ``spiked'', in that for $\delta k\sim 1$ around the $k_F$, we have $\delta |G|\sim 10^9$. We also consider the black hole with $r_0=3$, $r_1=1$ and $r_2=2$. In other words, all the roots are positive. We find a Fermi surface at $qQ=5.3042024$ and $k_F=0.264110954$.  In this case, the spiked maximum of $G_1$ is also in the order of $10^9$.

\begin{figure}[ht]
\center
\includegraphics[width=7cm]{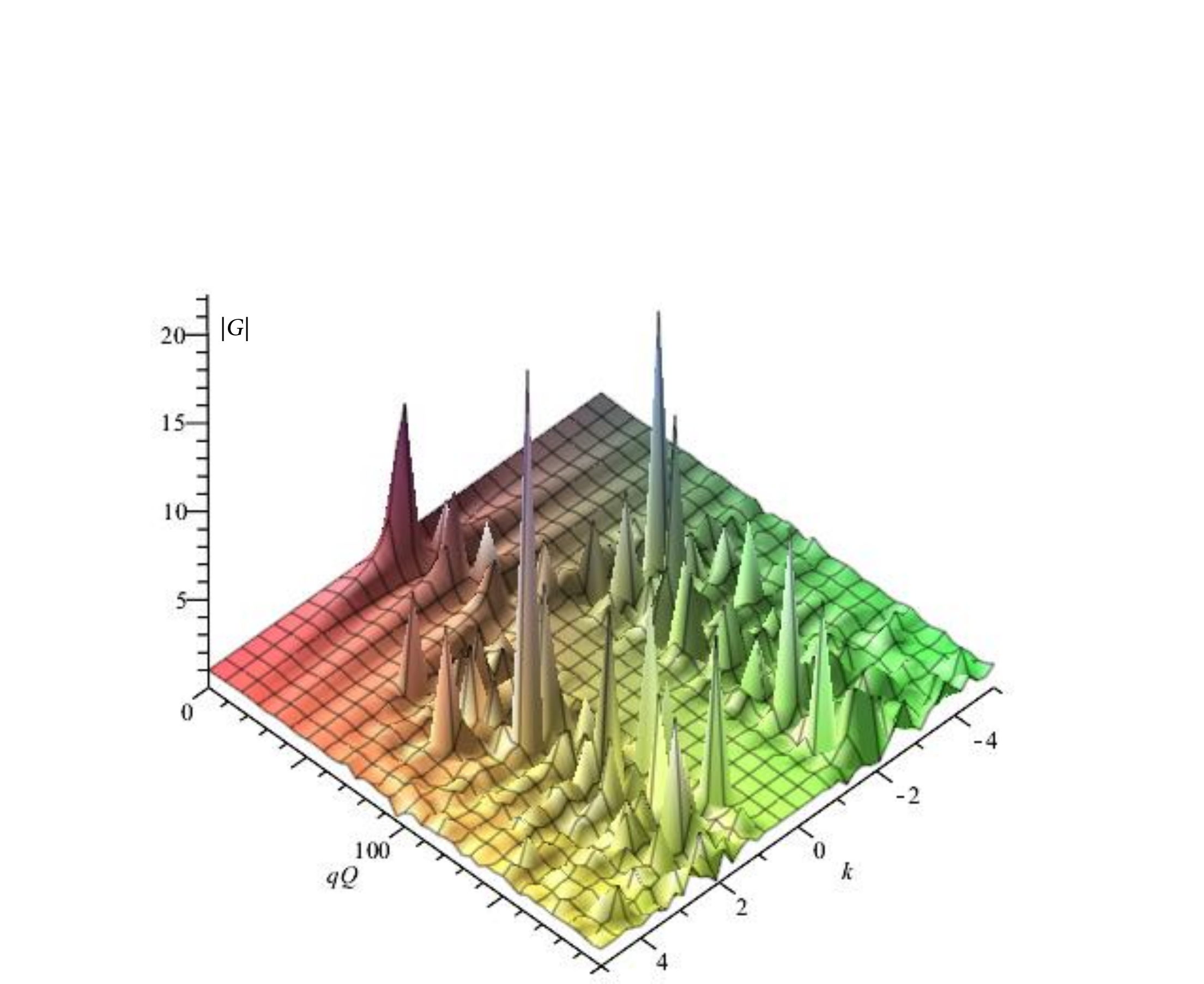}
\caption{The 3D plot of $|G|$ in the $(k,qQ)$-plane shows a fascinating rich structure of spiked maxima.  The black hole is specified by $r_0=3$, $r_1=1+{\rm i}$ and $r_2=1-{\rm i}$.}
\label{finalf}
\end{figure}

The final example is provided in Fig.~\ref{finalf} on the black hole $r_0=3$, $r_1=1+{\rm i}$ and $r_2=1-{\rm i}$. In this case, $f$ has only one real root, and hence there is no inner horizon.  Again, the 3D figure of $|G|$ in the $(k,qQ)$-plane shows a fascinating rich structure of spiked maxima.

To end this section, We are obliged to acknowledge that we have used extensively the properties of the general Heun's functions listed in \cite{heun1,heun2}.  For our purpose, we find that the following identities are particularly useful.

({\bf 1}) The Heun's function $H\ell(a,b; \alpha,\beta,\gamma,\delta; z)$ can be written as the linear combination of the following two Heun's functions:
\begin{eqnarray}
z^{-\alpha} H\!\ell\Big(\ft{1}{a}, \alpha (\epsilon - \beta) + \ft{\alpha}{a} (\delta - \beta) + \ft{b}{a}; \alpha, \alpha - \gamma + 1, \alpha -\beta + 1, \delta; \ft1{z}\Big),\cr
z^{-\beta} H\!\ell\Big(\ft{1}{a}, \beta (\epsilon - \beta) + \ft{\beta}{a} (\delta - \beta) + \ft{b}{a}; \beta, \beta - \gamma + 1, \beta -\alpha + 1, \delta; \ft1{z}\Big).\label{nonidentity}
\end{eqnarray}
Unfortunately, the precise coefficients are not yet known in the literature.  Owing to the lacking of this knowledge, we were unable to perform the subtle extremal limit of the wave solutions and the Green's function from those of most general non-extremal backgrounds.

({\bf 2}) The following is a derivative identity that we find useful to calculate $u_+$ from $u_-$:
\begin{equation}
\fft{d}{dz}H\!\ell(a,b;0,\beta, \gamma,\delta;z)=H\!\ell(a,b+ \gamma + (a-1)\delta + \beta+1;2,\beta+1,\gamma+1,\delta+1,z).
\end{equation}

({\bf 3}) We also used the following identity to put $u_+^{\rm in}$ in the form as the complex conjugate of $u_-^{\rm out}$:
\begin{eqnarray}
&&H\!\ell(a,b;\alpha,\beta,\gamma,\delta;z) = (1-z)^{1-\delta} (1-\ft{z}{a})^{-\alpha-\beta+\gamma +\delta} \times\cr
&&\qquad\qquad H\!\ell\Big(a,b-\gamma((\delta-1)a + \beta - \gamma - \delta); \gamma -\beta +1,\gamma -\alpha +1, \gamma, 2-\delta; z\Big).
\end{eqnarray}

\section{Conclusions}

In this paper, we have considered conformal gravity in four dimensions.  It is constructed from the Weyl-squared term together with the minimally-coupled Maxwell field. The most general spherically-symmetric charged AdS black holes up to an arbitrary conformal factor was constructed in \cite{Riegert:1984zz}.  We generalize the result to the torus $T^2$ and hyperbolic $H^2$ topologies.  We analyze the global structure and obtain the full set of thermodynamical quantities that satisfy the first law of thermodynamics.

Aside from the facts that conformal gravity plays an important role in critical gravity and also in ${\cal N}=1$, $D=4$ off-shell supergravity, the main reason that we construct the $T^2$-symmetric charged AdS black holes is to use them as gravitational backgrounds to study the properties of the boundary field theory.  Charged AdS black holes such as the extremal RN black holes in general dimensions were demonstrated to be dual to Fermi and non-Fermi liquids and Fermi surfaces were determined.  In the literature, the analytic Green's functions for general $(\omega,k)$ are lacking and properties of $G(\omega,k)$ were obtained through numerical analysis. In this paper, we have demonstrated that the Dirac equation for a charged massless spinor on the AdS black holes in conformal gravity can be solved exactly.  This allows us to obtain the exact Green's function for all $(\omega,k)$.  Since in conformal gravity the black holes are given up to arbitrary conformal transformations, this leads to the question how general our results are.  We show that the Green's function, and hence the Fermi surfaces are invariant under the conformal transformations that preserve the $T^2$ isometry. Thus our results cover all the $T^2$-symmetric black holes. Some salient results of this paper were reported earlier in \cite{lwshort}.

The Green's function from the general black hole is expressed in terms of the general Heun's functions.  The Heun's functions are much less studied and there is no package in Mathematica and only a limited package in Maple.  This makes it difficult to analyse the Fermi surfaces and their properties in detail using these programs. Nevertheless, we made numerical plots of $G(0,k)$, which show fascinating rich structures of spiked maxima in the plane spanned by $(k,q)$.

The most general black hole solution contains three non-trivial parameters. We consider some special subclasses of the black holes with two non-trivial parameters: the horizon radius and the charge. The solution can be viewed as being pseudo-supersymmetric from the point of view ${\cal N}=1$, $D=4$ off-shell supergravity and was obtained in \cite{luwangsusylif}. We find that the Heun's functions are reduced to hypergeometric functions.  The Green's function for general $(\omega,k)$ becomes fully analysable.  The Fermi surfaces are defined by the poles of $G(0, k)$. In the extremal limit of this subclass of backgrounds, we find that there can only be one Fermi surface, corresponding to having $\nu=0$.  We obtain the small $\omega$ behaviour near the Fermi surface.  When the background is non-extremal, there is no Fermi surface at all for real $\nu$.  However, Fermi surfaces emerge when $\nu$ is pure imaginary.

It should be emphasized that qualitative features of the charged black holes in conformal gravity are similar to the RN black holes.  Although black holes in conformal gravity can have slower falloffs associated with the massive spin-2 hair, the asymptotic spacetime is nevertheless AdS$_4$, the same as the RN black hole.  Furthermore, the near-horizon geometry in the extremal limit contains an AdS$_2$ factor, the same as the RN black holes.  Thus our results are qualitatively similar to those of RN black holes.  In the extremal and near-extremal limits, we indeed reproduced the qualitative features obtained in the extremal or near extremal RN black holes for small $\omega$. Also as in the RN black holes, AdS$_2$ metal can arise for small enough $q$ for which no Fermi surface emerges. Of course, our analytical results allow one to study the Green's function for larger $\omega$, and we expect that these qualitative features may also apply in the RN black holes.

Our explicit results also demonstrate that there are qualitative differences between the extremal background whose decoupling limit is AdS$_2\times T^2$ and the non-extremal background whose topology is $\mathbb{R}^2\times T^2$.  For both extremal and non-extremal backgrounds, if we have wave solutions for generic $(\omega, k)$, we can organise the general solution in terms of in-falling and outgoing modes on the horizon.  The boundary condition of a black hole horizon selects the in-falling solution, which is associated with the retarded Green's function.  Since the Fermi surfaces are determined by the Green's function with $\omega=0$.  It is thus tempting to think that we only need solve for wave functions with $\omega=0$, in which case, the equations tend to be reduced significantly.  However, once $\omega$ is set to zero, we cannot organise the solutions in terms of the in-falling and outgoing modes.  In the extremal background, for real $\nu$, the boundary condition on the horizon could still rule out one of the two modes, since it diverges on the horizon.  Furthermore, if $\nu$ is imaginary, the modes become oscillatory in the near-horizon region.  This signals instability of the system.  In fact, as demonstrated in \cite{flmv}, in the extremal or near-extremal limits with AdS$_2\times T^2$ horizon, the Green's function for small $\omega$ can always be determined as long as $G(0,k)$ is known.

The situation is quite different in the general non-extremal case.  When $\omega$ is set to zero, we find that both modes are finite on the horizon.  Thus for non-extremal backgrounds, there is no known procedure of selecting the correct mode if we have only solutions with $\omega=0$.  This implies that in order to obtain the Fermi surfaces, we have to obtain solutions with non-vanishing $\omega$, even though the Fermi surfaces are determined by the properties of the Green's function with vanishing $\omega$. Another difference is that the in-falling wave solution with $\omega=0$ is not oscillatory even when $\nu$ is imaginary.  Thus there is no instability associated with imaginary $\nu$. Using our explicit example, we show that for imaginary $\nu$, the number of Fermi surfaces increases as the black hole approaches the extremality, and the number becomes infinite in the extremal limit. This provides an explanation of the origin of the instability associated with imaginary $\nu$ in the extremal backgrounds.

We also study the Green's function for the general extremal solution with two non-trivial parameters, characterised also by the horizon radius and the charge.  The Green's function $G(\omega,k)$ is now expressed in terms of confluent Heun's functions.  However, $G(0,k)$ is given in some simpler hypergeometric functions.  We find multiple numbers of Fermi surfaces arising in this case. These Fermi surfaces can be ones for either Fermi liquids $(\nu>1/2)$ or non-Fermi liquids $(\nu<1/2)$, including the extreme case $\nu=0$.

Our further analysis on the wave solutions and the Green's function for the most general black hole indicate the $\nu$ parameter that appears in the non-extremal ``BPS'' solution is purely accidental. In the Green's function and the wave solutions in the most general black hole, there is no parameter combination that resembles $\nu$.  The reality condition on $\nu$ for the extremal background which implies $k\ge k_{\rm min}(q)$ is relaxed in the non-extremal case where $(k,q)$ can take all real values.  The numerical plots of the Green's function in the $(k,q)$ plane show a fascinating rich structure of spiked maxima.

In this paper we focus on the study of the Green's function on real frequency $\omega$.  It is also of great interest and physically relevant to study the Green's function on the complex $\omega$-plane.  For the simpler ``BPS'' extremal black holes, this was done in \cite{lwshort}. The contour plots of the Green's function on the complex $\omega$-plane were analysed, revealing a rich structure of zeros and poles.  From the motions of the poles, we can determine a dispersion relation that fits the data for a large $(\omega,k)$ region. The contour plots reveal that for the extremal black holes, the poles of the Green's function associated with the Fermi surfaces lie exactly on the real axis of the complex $\omega$-plane, whilst they lie below the real axis for non-extremal black holes.

Conformal gravity that involves the Maxwell field is uniquely four dimensional and the dual field theories have $(2+1)$ dimensions. Although the exact conformal field theory that is dual to conformal gravity on AdS is still unknown, our results show that conformal gravity and its charged black holes provide useful playgrounds for investigating properties in both Fermi and non-Fermi liquids on a two-dimensional plane, not only at zero temperature but also at any finite temperature.  The fact that the Dirac equation for a charged massless fermion is invariant under the $T^2$-preserving conformal transformations implies that conformal gravity is particularly suitable for studying such fermionic system.

\section*{Acknowledgement}

H.L. is grateful to KIAS for hospitality during part of the course of this work. J.L. is supported in part by NSFC grants 11031005, 11135006 and 11125523.
The research of H.L. is supported in part by NSFC grants 11175269 and 11235003.
The research of Z.L.W. is supported in part by the National
Research Foundation of Korea (NRF) funded by the Ministry of Education, Science
and Technology with the grant number 2010-0013526.

\end{document}